\documentclass[a4paper]{jpconf}
\usepackage{graphicx}
\usepackage{amsmath}
\usepackage{xcolor}
\usepackage{amsfonts,amsmath,amssymb, bm}
\usepackage{upgreek,xspace}

\newcommand{\eqlab}[1]{\label{eq:#1}}
\renewcommand{\eqref}[1]{Eq.~(\ref{eq:#1})}
\newcommand{\eqsref}[2]{Eqs.~(\ref{eq:#1}) and~(\ref{eq:#2})}

\newcommand{\seclab}[1]{\label{sec:#1}}
\newcommand{\secref}[1]{Section~\ref{sec:#1}}

\newcommand{\appref}[1]{\ref{app:#1}}

\newcommand{\tabref}[1]{Table~\ref{tab:#1}}

\newcommand{\figref}[1]{Fig.~\ref{fig:#1}}

\newcommand{\figlab}[1]{\label{fig:#1}}
\newcommand{\equal}{\!=\!}
\newcommand{\minus}{\!-\!}
\newcommand{\plus}{\!+\!}

\newcommand{\vsubsup}[3]{\bm{#1}_{#2}^{\scriptscriptstyle #3}}

\newcommand{\vsup}[2]{\bm{#1}^{\scriptscriptstyle #2}}
\newcommand{\ssub}[2]{{#1}_{#2}}

\newcommand{\omSub}[1]{\omega_{ #1 }}
\newcommand{\OmSub}[1]{\Omega_{ #1 }}
\newcommand{\lamSub}[1]{\lambda_{ #1 }}
\newcommand{\delSub}[1]{\delta_{ #1 }}
\newcommand{\delsub}[1]{\delta_{\scriptscriptstyle #1 }}
\newcommand{\gamSub}[1]{\gamma_{ #1 }}
\newcommand{\gamsub}[1]{\gamma_{\scriptscriptstyle #1 }}
\newcommand{\GamSub}[1]{\Gamma_{ #1 }}

\newcommand{\lamFSR}{\lambda_{\scriptscriptstyle \text{FSR}}}
\newcommand{\OmFSR}{\Omega_{\scriptscriptstyle \text{FSR}}}
\newcommand{\omout}{\omega_{ \text{out}}}
\newcommand{\hatdsub}[2]{\hat{#1}^{\dagger}_{ #2}}
\newcommand{\hatsub}[2]{\hat{#1}_{#2}}

\newcommand{\So}[1]{S_{\text{out}, #1}}
\newcommand{\Si}[1]{S_{\text{in}, #1}}
\newcommand{\dotSi}[1]{\dot{S}_{\text{in}, #1}}

\newcommand{\omref}{\omega_{\text{ref} }}
\newcommand{\braket}[1]{\langle #1 \rangle}

\newcommand{\sti}[1]{\tilde{s}_{\text{in}, #1}}
\newcommand{\Sto}[1]{\tilde{S}_{\text{out}, #1}}
\newcommand{\Sti}[1]{\tilde{S}_{\text{in}, #1}}

\begin{document}


\title{Unidirectional frequency conversion in microring resonators for on-chip frequency-multiplexed single-photon sources}

\author{Mikkel Heuck$^{1,*}$, Jacob Gade Koefoed$^{1}$, Jesper Bjerge Christensen$^{1}$, Yunhong Ding$^{1}$, Lars Hagedorn Frandsen$^{1}$, Karsten Rottwitt$^{1}$, Leif Katsuo Oxenl\o we$^{1}$}
\address{${}^{1}$Department of Photonics Engineering, Technical University of Denmark, Building 343, 2800 Kgs. Lyngby, Denmark}
\ead{${}^{*}$mrheuck@gmail.com}

\begin{abstract}
Microring resonators are attractive for low-power frequency conversion via Bragg-scattering four-wave-mixing due to their comb-like resonance spectrum. However, conversion efficiency is limited to 50\% due to the equal probability of up- and down-conversion. Here, we demonstrate how two coupled microrings enable highly directional conversion between the spectral modes of one of the rings. An extinction between up- and down-conversion of more than 40$\,$dB is experimentally observed. Based on this method, we propose a design for on-chip multiplexed single-photon sources that allow localized frequency modes to be converted into propagating continuous-mode photon wave packets using a single operation. The key is that frequency conversion works as a switch on both spatial and spectral degrees of freedom of photons if the microring is interferometrically coupled to a bus waveguide. Our numerical results show 99\% conversion efficiency into a propagating mode with a wave packet having a 90\% overlap with a Gaussian for a ratio between intrinsic and coupling quality factors of 400.
\end{abstract}


\section{Introduction}

Single photon sources based on probabilistic parametric processes, such as down-conversion or four-wave-mixing (FWM), can be made near-deterministic by use of multiplexing. 
Quantum frequency conversion was recently demonstrated as a powerful tool to enable multiplexing of individual photons generated with different frequencies~\cite{Puigibert2017, Joshi2018}. The main advantage of using the spectral degree of freedom of photons is that all modes can be combined in a single device, such as a nonlinear crystal~\cite{Puigibert2017}, fiber~\cite{Joshi2018}, or resonator~\cite{Li2016}. This is generally not the case for spatial- or temporal degrees of freedom~\cite{Collins2013, Kaneda2015b, Bonneau2015, Mendoza2016} where the number of devices (and therefore the total loss) increases with the number of multiplexed modes. Additionally, noise-free conversion is possible between frequency modes using Bragg-scattering four-wave-mixing (BS-FWM)~\cite{McKinstrie2005,Vernon2016}. So far, all demonstrations of multiplexing were implemented using fiber or free-space optics due to the difficulty of achieving on-chip quantum feedback control. 
\begin{figure}[!t]
  \centering
  \hspace{-2mm}
  \includegraphics[height=4.5cm]{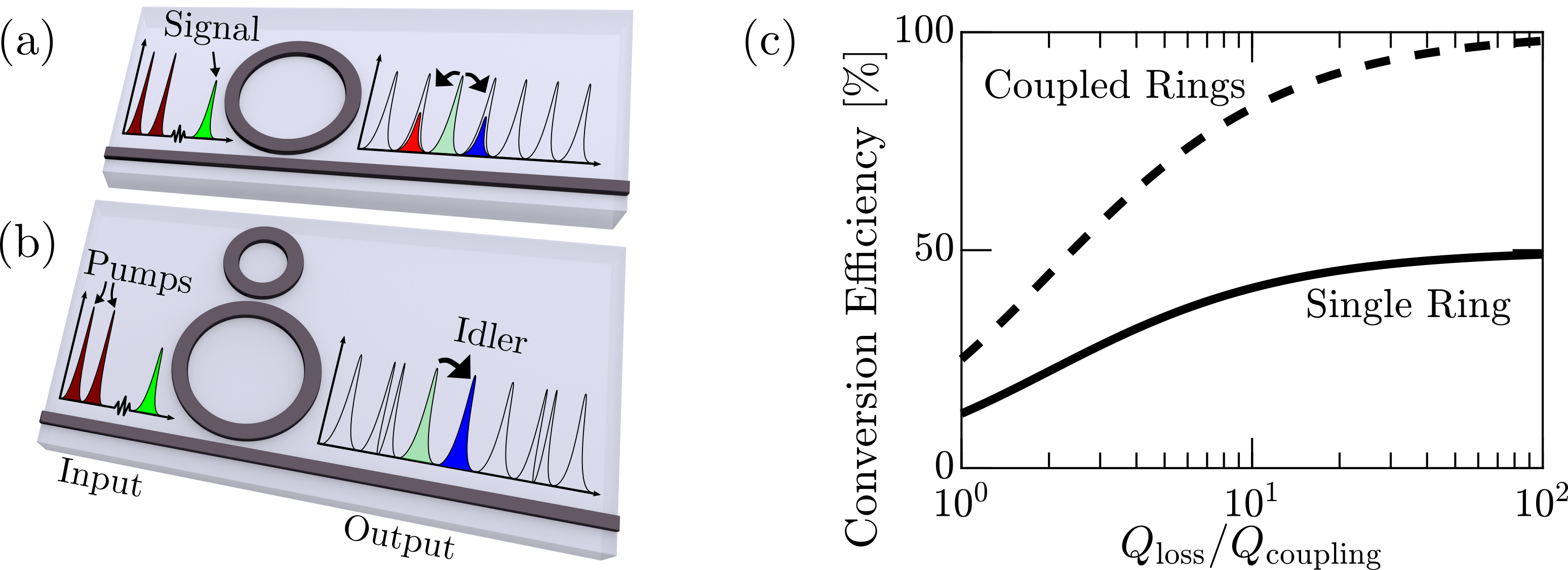}
 \caption{(a) Illustration of Bragg-scattering FWM in a single ring resonator. Two strong pump fields are injected into the ring along with a signal field. In the FWM process, a photon is exchanged between the pumps while a signal photon is converted in energy by an amount corresponding to the energy difference between the pumps. The comb-like mode spectrum of the ring resonantly enhances both up- and down-conversion because those processes only differ by reversing the roles of the pumps as donor or acceptor of a photon. (b) Mode splitting induced by coupling between two ring resonators enables unidirectional frequency conversion by only providing resonant enhancement to the up-conversion process. (c) Calculated up-conversion efficiency by considering the signal mode and its nearest neighbors as well as one mode of the smaller ring (see~\appref{unidirectional theory} for details).}
\figlab{concept figure}
\vspace{-3mm}
\end{figure}
However, compact and energy efficient sources suitable for large-scale quantum information processing require devices based on photonic integrated circuits (PICs). Ring resonators are natural candidates owing to their comb-like mode spectrum (see~\figref{concept figure}a), which has been used to demonstrate frequency conversion~\cite{Li2016} and high-dimensional entanglement~\cite{Kues2017}. Dispersion engineering enables negligible variation (relative to the resonance linewidth) of the free-spectral-range (FSR) over a large bandwidth. Photon pair generation by spontaneous-four-wave-mixing (SFWM) is therefore resonantly enhanced across a very large number of modes~\cite{Okawachi2011}, which may be multiplexed using BS-FWM frequency conversion. However, the conversion efficiency between two modes is limited to 50\% due to the symmetry between up- and down- conversion, as illustrated in Figs. \ref{fig:concept figure}a,c.

The maximum 

In this work, we demonstrate a PIC device that uses mode-coupling to allow frequency conversion between resonances with an extinction ratio above 40 dB. The concept (illustrated in~\figref{concept figure}b) is based on coupling two rings where the FSR of one is an integer multiple of the other. When their resonances align, the coupling-induced mode-splitting effectively eliminates either the up- or down-converted resonance leading to near-unity conversion efficiency (see Figs. \ref{fig:concept figure}b,c). Coupled rings have previously been used for dispersion engineering in FWM applications~\cite{Gentry2014,Xue2015}. Based on our demonstration of unidirectional frequency conversion, we propose a PIC single-photon source that multiplexes the dense spectral modes of microring resonators to enable quasi-deterministic emission. Its multiplexing protocol only involves switching the classical pump fields $-$ making the single-photon loss independent of the number of multiplexing modes~\cite{Joshi2018}. The device is very compact since photon creation and frequency conversion occur in the same resonator structure. The output photons have very high spectral purity~\cite{Vernon2017}, which is essential for high-visibility multi-photon interference. Additionally, we show how the temporal wave packet of emitted photons may be controlled by shaping the BS-FWM pump fields.\\


This article is organized as follows: In~\secref{experiment} we present our experimental results demonstrating unidirectional frequency conversion between modes of a ring resonator.~\secref{new protocol} presents our proposal for using this concept in a PIC implementation of a frequency-multiplexed single-photon source. We conclude in~\secref{discussion} with a discussion of the challenges involved with realizing our proposal.

\section{Experiment}\label{sec:experiment}
To demonstrate unidirectional frequency conversion, we fabricated a 
\begin{figure}[!h]
  \centering
  \includegraphics[height=4.0cm]{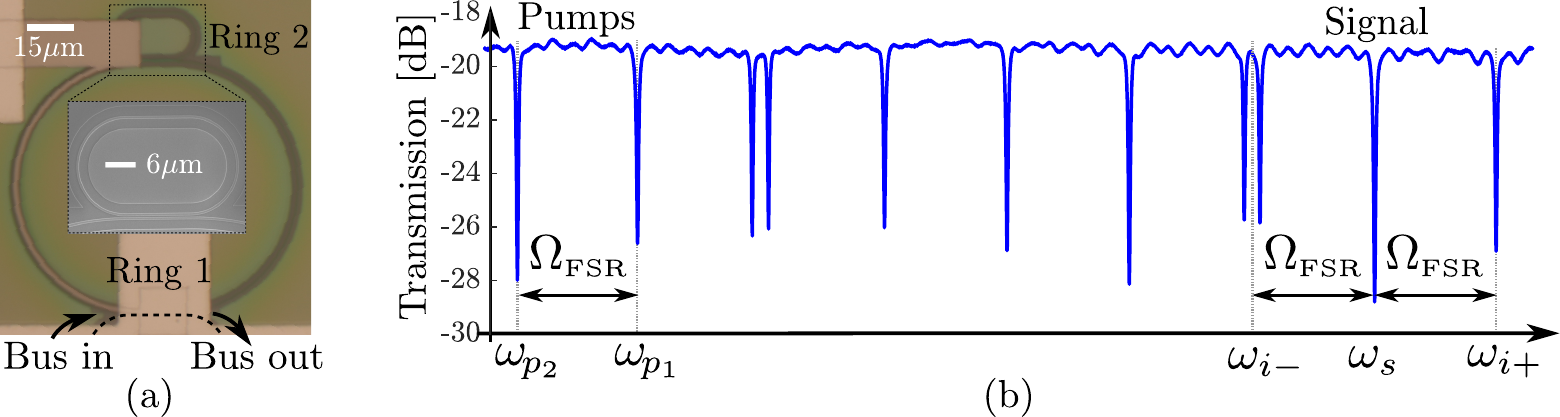}
 \caption{(a) Optical microscope image of fabricated double-ring device. A heater covering the left side of ring 1 is used to modify its resonance frequencies. The inset shows a scanning electron microscope image of ring 2. (b) Measured transmission spectrum showing which modes are used for the signal ($\ssub{\omega}{s}$) and two pumps ($\ssub{\omega}{p_1}$ and $\ssub{\omega}{p_2}$) in the BS-FWM experiment. }
\figlab{intro fig}
\end{figure}
device (shown in~\figref{intro fig}a) consisting of two coupled microring resonators realized in a silicon-on-insulator (SOI) material (250 nm silicon on a 3$\,\upmu$m thick buried oxide layer) using electron-beam lithography. The waveguides are 500$\,$nm wide and the circumference of ring 1 (2) is 324$\,\upmu$m (81$\,\upmu$m) such that the FSR of ring 1, $\OmFSR$, is four times smaller than that of ring 2. The device is covered by a 1$\,\upmu$m thick oxide layer on top of which a heating element is formed by a thin titanium wire, see~\figref{intro fig}a. Grating couplers~\cite{Ding2013a} are used for coupling in and out of the chip. 

In the BS-FWM experiment we pump ring 1 on the two resonances at $\omSub{p_1}$ and $\omSub{p_2}$ (see~\figref{intro fig}b) using continuous wave (CW) lasers. A weak CW laser is used for the signal and its wavelength is scanned across the resonance at $\omSub{s}$. For each wavelength, $\lambda$, of the signal laser, the spectrum of the generated idler fields are measured in the vicinity of the up- and down-converted modes at $\omega_{i+}$ and $\omega_{i-}$, respectively. This gives rise to two-dimensional idler power maps. Our setup for the BS-FWM experiment is shown in~\figref{setup}. 
\begin{figure}[!h]
  \centering
  \includegraphics[height=3.3cm]{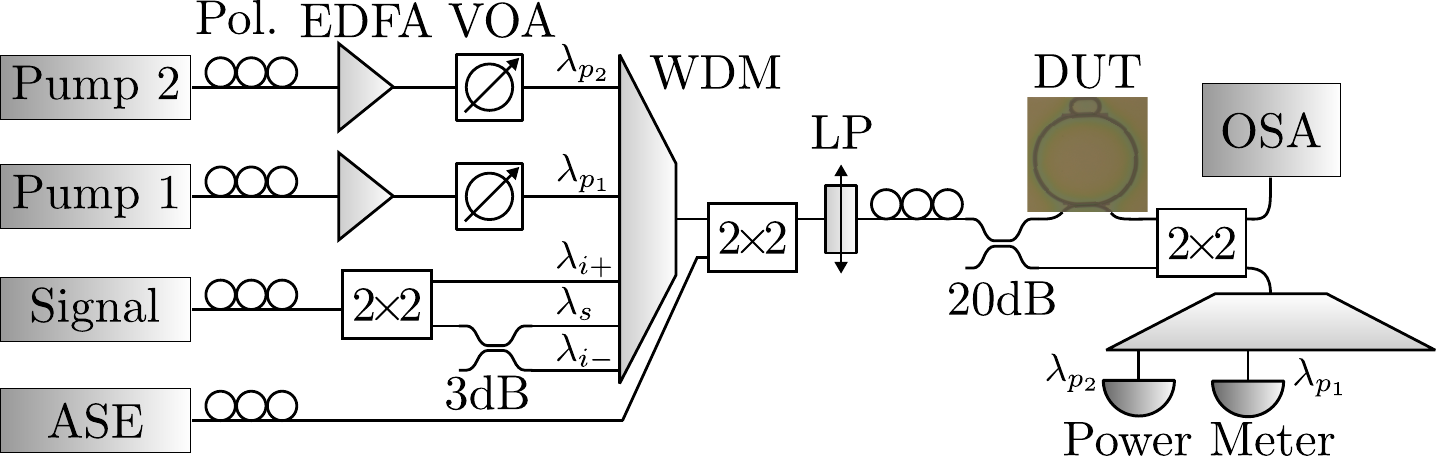}
 \caption{Experimental setup. Pol: Polarization controller, EDFA: Erbium-doped fiber amplifier, VOA:Variable optical attenuator, WDM: Wavelength division multiplexing filter,  LP: Linear polarizer, DUT: Device under test, OSA: Optical spectrum analyzer, ASE: Amplified spontaneous emission source.}
\figlab{setup}
\end{figure}
The three CW lasers (pump 1, pump 2, and signal) are combined using a WDM filter and both input and output can be sent to an OSA using a two-by-two switch. Additionally, the pump output power can be monitored by power meters while tuning their wavelengths to thermally lock them to the ring resonances~\cite{Li2016}. 
A broadband ASE source is used to measure linear transmission spectra. 

The heater on ring 1 is used to align the resonances of the two rings.~\figref{BSFWM}a shows the measured transmission as a function of wavelength (close to $\lamSub{i-}$) and heater voltage. The spectrum exhibits an avoided crossing typical of strongly coupled systems~\cite{Reithmaier2004}.
\begin{figure}[!h]
  \centering
  \includegraphics[height=6.5cm]{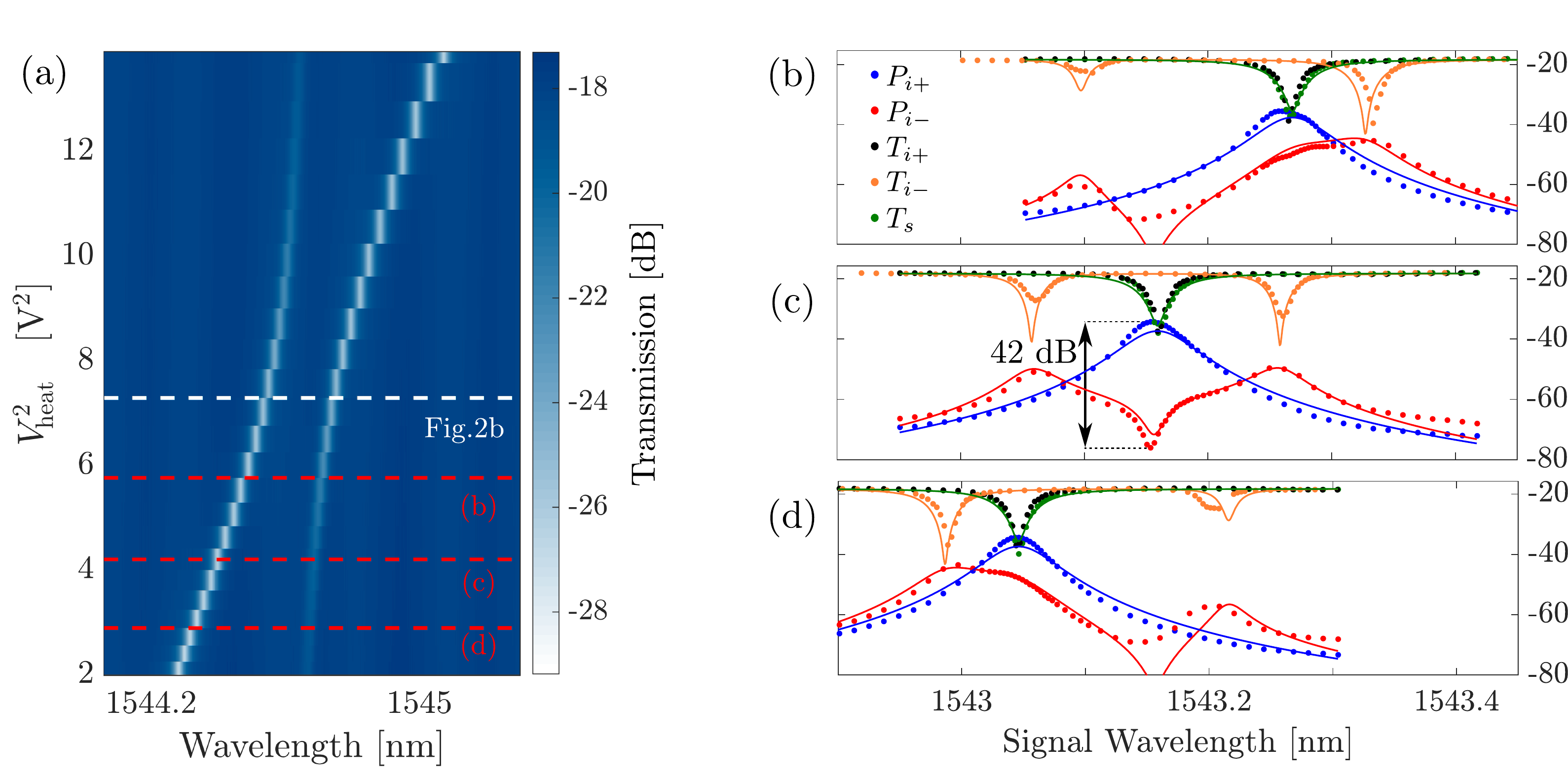}
 \caption{(a) Transmission as a function of  wavelength and heater voltage in the vicinity of $\lamSub{i-}$. (b)-(d) Idler output power as a function of the signal laser wavelength, $\lambda$,  from the down-converted resonance (red) and up-converted resonance (blue) as well as the transmission near $\lamSub{i-}$ (orange), $\lamSub{s}$ (green), and $\lamSub{i+}$ (black). Solid lines are fits using~\eqsref{trans CW CMT}{eta CW} with the parameters listed in~\tabref{device params}. }
\figlab{BSFWM}
\end{figure}
From the two-dimensional idler power maps, the diagonal cross sections corresponding to the power emitted near the up-converted mode, $P_{i+}(\lambda, \lambda \minus \lamFSR$), and down-converted mode, $P_{i-}(\lambda, \lambda \plus \lamFSR$), are plotted in Figs.~\ref{fig:BSFWM}b-d. The corresponding heater voltages are indicated in~\figref{BSFWM}a. The three plots correspond to the resonance of ring 1 being red-detuned (b), aligned (c), and blue-detuned (d) with respect to the resonance of ring 2. Note that the symmetric case in~\figref{BSFWM}c requires a lower heater voltage compared to~\figref{intro fig}b due to the thermal red-shift induced by the pumps. Figs.~\ref{fig:BSFWM}b-d also show transmission spectra of the down-converted mode, $T_{i-}(\lambda\plus\lamFSR)$, signal mode, $T_s(\lambda)$, and up-converted mode, $T_{i+}(\lambda\minus\lamFSR)$, while the pumps are on. In~\appref{CMM}, we calculate the transmission near the three modes
\begin{subequations}\eqlab{trans CW CMT}
\begin{align}
T_{i+}(\Omega)  &= T_{s}(\Omega)  =T_{\text{cpl}}^2 \Bigg| 1-\frac{\gamSub{}}{    \frac{\gamma + \overline{\gamma}_{L_1}}{2} - i\Omega } \Bigg|^2 \eqlab{trans CW CMT i+},\\
T_{i-}(\Omega)  &= T_{\text{cpl}}^2 \Bigg| 1-\frac{\gamSub{}}{ \frac{\gamSub{}+\overline{\gamma}_{L_1}}{2} - i\Omega + \frac{g^2}{\gamSub{L_2}/2 - i(\Omega-\Delta_{ab}) }} \Bigg|^2. \eqlab{trans CW CMT i-}
\end{align}
\end{subequations}
The cavity-waveguide coupling rate, $\gamma$, is assumed equal for all three modes and $\Omega$ is the detuning from each of the three resonances. The loss rate of ring 2 is $\gamSub{L_2}$ and in ring 1 it is $\overline{\gamma}_{L_1}\equal \gamSub{L_1}\plus \gamsub{\rm{FCA}}$, where $\gamsub{\rm{FCA}}$ is the loss rate from free carrier absorption due to the presence of the BS-FWM pumps. The detuning between the resonances of the rings near the down-converted mode at $\lamSub{i-}$ is $\Delta_{ab}\equal \omSub{b}-\omSub{i-} \plus \delsub{\rm{NL}}$, where $\omSub{b}$ is the resonance of ring 2 and $\delsub{\rm{NL}}$ is a nonlinear shift caused by heating from the BS-FWM pumps. The coupling rate between ring 1 and 2 is $g$ and $T_{\text{cpl}}$ is the coupling loss of one grating coupler. The up- and down conversion efficiency is (see~\appref{CMM} for derivations)
\begin{align}\eqlab{eta CW}
	\eta_{i+} &= \left| \frac{4 \gamma \bar{\chi}}{(\GamSub{1} \minus i2\Omega)^2 \plus 4\bar{\chi}^2\big(1 \plus \zeta^{-1}\big)} \right|^2 \!\!,~~~~\eta_{i-} = \frac{1}{|\zeta|^2}\eta_{i+}.
\end{align} 
The total decay rate is $\GamSub{1}\equal \gamSub{} \plus \overline{\gamma}_{L_1} $ and the extinction ratio, $\zeta$, is given by
\begin{align}\eqlab{extinction ratio}
\zeta(\Omega)  &= \frac{4g^2 }{\big[\gamSub{L_2}+i2(\Delta_{ab}-\Omega)\big]\big(\GamSub{1} -i2\Omega\big)} + 1 .
\end{align}
The nonlinear coupling rate due to the BS-FWM pumps is~\cite{Vernon2016}
\begin{align}\eqlab{chi estimate 2}
\bar{\chi} =\frac{8  \omega c n_2}{n_{\rm{eff} }'^{2} V_{\rm{ring}}}\frac{\gamSub{}}{\GamSub{1}^2}\sqrt{P_{\!p_1} P_{\!p_2}},
\end{align}
where $n_2$ is the nonlinear index, $c$ the speed of light, $n_{\rm{eff} }'$ the real part of the mode index of the waveguide, $V_{\rm{ring}}$ the volume of ring 1, and $P_{p_1}$ and $P_{p_2}$ the input pump power in the waveguide immediately before the ring. The maximum up-conversion efficiency is
\begin{align}\eqlab{eta CW max}
\eta_{i+}^{\rm{max}} =  \frac{1 + 4G^2}{2 + 4G^2} \left(\frac{\gamSub{}}{\GamSub{1}}\right)^2 ,
\end{align} 
where $G\equiv g/\sqrt{\overline{\gamma}_{L_1}\GamSub{1}}$.~\eqref{eta CW max} shows that $\eta_{i+}$ is bounded by 50\% when $G\equal 0$ (corresponding to a single ring), but approaches unity as $G\rightarrow\infty$ and $\GamSub{1}\rightarrow\gamma$.\\

In~\appref{model experiment comparison}, we estimate the values of parameters in the model using a step-wise fitting procedure where the measured data is compared to~\eqsref{trans CW CMT}{eta CW} both with- and without the pumps. The results are listed in~\tabref{device params}. The solid lines in Figs.~\ref{fig:BSFWM}b-d plots~\eqsref{trans CW CMT}{eta CW} using the fitted parameter values and shows a good agreement with the measured data. 
\renewcommand{\arraystretch}{1.3}
\begin{table}[htbp]
\centering
\begin{tabular}{p{0.4cm} p{0.06cm} p{1.5cm}  p{0.05cm} | p{0.05cm}   p{0.4cm}  p{0.06cm} p{2.75cm}  }
\hline
 \multicolumn{3}{c}{Measured } & \multicolumn{2}{c}{ } & \multicolumn{3}{c}{Fitted}\\
 \hline
$T_{\rm{cpl}}$  &=& $-9.4\;\text{dB}$ 		& && $\gamma$  	&=& $27.4 \times 10^{9} \,\mathrm{rad/s} $ \\
$P_{p_1}$  		&=& $3.9\;\text{dBm}$ 		& && $\gamSub{L_1}$ &=& $10.9 \times 10^{9} \,\mathrm{rad/s} $ \\
$P_{p_2}$  		&=& $4.2\;\text{dBm}$ 		& && $\gamSub{L_2}$ &=& $8.02 \times 10^{9} \,\mathrm{rad/s}$ \\
$P_{s}$  		&=& $-11.6\;\text{dBm}$ 	& && $g$  &=& $78.5 \times 10^{9} \,\mathrm{rad/s}$ \\
$\eta_{+}$  	&=& $-22.6\;\text{dB}$ 		& && $\gamSub{\scriptscriptstyle \text{FCA}}$  	&=& $10.0 \times 10^9 \,\mathrm{rad/s}$ \\ 
$|\zeta|^2$ &=& $41.7\;\text{dB}$ 			& && $ \bar{\chi}$  			& =& $1.09 \times 10^{9} \,\mathrm{rad/s}$\\
\hline
\end{tabular}
\caption{Device parameters. The in- and out-coupling, $T_{\rm{cpl}}$, is estimated by assuming the transmission away from any resonance is $T_{\rm{cpl}}^2$. The power in the bus waveguide immediately before the ring is estimated as $P_{n}\equal T_{\rm{cpl}} P_{\text{in},n}$, where $P_{\text{in},n}$ is the power measured before the fiber array. The conversion efficiency is estimated from the ratio of the peak of the up-converted power, $P_{i+}$, to the input power, $P_s$.  }
  \label{tab:device params}
\end{table}
%
Our measured conversion efficiency is 0.55\% (-22.6$\,$dB) and limited by pump power and two-photon absorption (TPA). Reaching near-unity efficiency requires a material without TPA~\cite{Li2016}. The estimated nonlinearity based  on~\eqref{chi estimate 2} is $1.68\!\times\!10^9\,$rad/s, which is reasonably close to the fitted value of $1.09\!\times\!10^9\,$rad/s (see~\tabref{device params}). 

Our main experimental result is a very high extinction between up- and down-conversion of more than 40\, dB when the ring resonances are aligned (see~\figref{BSFWM}c).

\section{On-Chip Frequency-Multiplexing Device}\label{sec:new protocol}
Having demonstrated a device concept enabling near-unity  frequency conversion efficiency between ring resonator modes allows us to design spectrally multiplexed single-photon sources for implementation in PICs. In order to both generate and frequency convert photons in the same resonator, control over the cavity-waveguide coupling for different modes is required. Interferometric coupling~\cite{Madsen1999} may be used to generate photon pairs where one, the signal photon, is decoupled from the bus waveguide whereas the other, the idler photon, is strongly coupled. Then, the signal photon remains in the resonator while the idler is routed to a detector that controls switches allowing specific BS-FWM pumps to convert the signal photon. A major advantage of our proposal is the elimination of spatial switches used in e.g. Ref.~\cite{Heuck2018} by converting signal photons to a common output mode, $\omout$, which is strongly coupled to the bus waveguide. It is even possible to shape the wave packet of the output photons by tailoring the temporal shape of the BS-FWM pumps,
\begin{figure}[!h]
  \centering
  	\includegraphics[height=5.5cm]{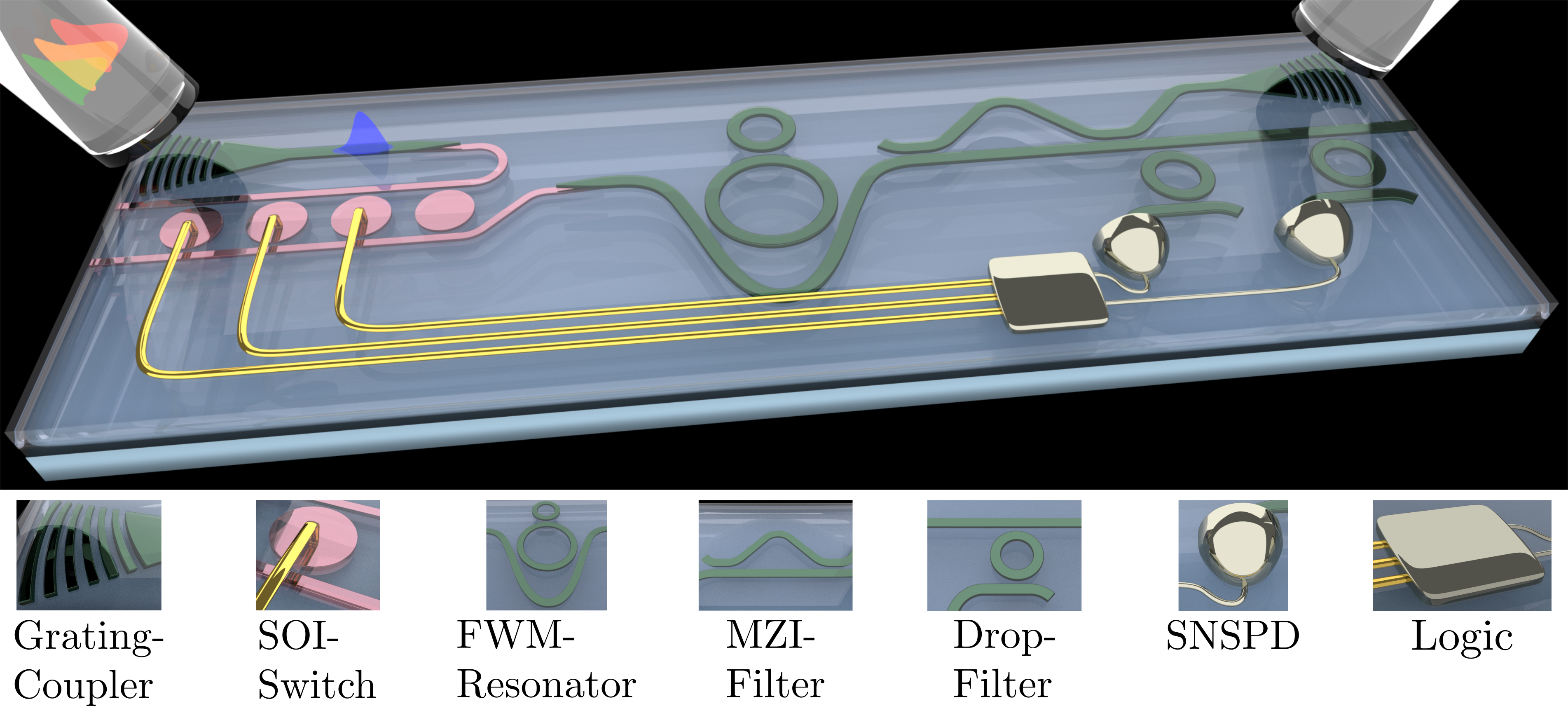}
 \caption{Schematic illustration of our proposed PIC.  The green layer is silicon nitride, pink is silicon, gold is superconducting material, and yellow is electrical wiring.}
\figlab{PIC}
\end{figure}
which we show in~\secref{frequency conversion}.

Our proposed PIC implementation is illustrated in~\figref{PIC}. 
It is based on a scalable multilayer silicon nitride (SiN) on SOI platform~\cite{Sacher2015, Shainline2017} that enables individual photons to be produced, frequency converted, and routed in the SiN layer with low loss and no TPA while switching of the pump fields occurs in the silicon layer~\cite{Gehl2017, Timurdogan2017}. Photon pair generation by SFWM and frequency conversion by BS-FWM occur in the FWM-resonator. It resembles the device in~\figref{intro fig}a except for its interferometric coupling, which is achieved by forming a Mach-Zehnder interferometer (MZI) between the bus waveguide and part of ring 1~\cite{Madsen1999}. Frequency-selective ring resonator drop-filters~\cite{Gentry2018} connect each idler mode to a specific superconducting-nanowire-single-photon-detector (SNSPD). The electrical signal from the detector is processed~\cite{McCaughan2014} and used to flip switches~\cite{Gehl2017,Timurdogan2017} controlling the passage of the BS-FWM pump fields. The rightmost SOI switch in~\figref{PIC} controls the passage of the SFWM pump at $\omSub{p}$ (blue pulse), its neighbor controls the common BS-FWM pump at $\omSub{p_2}$ (green pulse), while the rest control BS-FWM pumps at $\omSub{p_1,n}$ (yellow and red pulses) for each spectral multiplexing mode, $n\!\in\!\mathbb{Z}$. Note that the illustration in~\figref{PIC} is an example using $N\equal 2$ spectral modes.

The emission protocol consists of the following two steps: First, a SFWM pump pulse at $\omSub{p}$ generates photons at any of the mode pairs $\omSub{s,n}$ and $\omSub{i,n}$, where $s$ ($i$) means signal (idler) and $n\!\in\!\mathbb{Z}$ enumerates the spectral multiplexing modes. Signal photons remain in the resonator while idler photons are routed to the SNSPDs. Second, an idler detection causes two SOI switches~\cite{Gehl2017,Timurdogan2017} to flip allowing BS-FWM pumps at $\omSub{p_2}$ and $\omSub{p_1,n}$ (with $\omSub{p_2} \minus \omSub{p_1,n} \equal \omout \minus \omSub{s,n}$) to enter the resonator and frequency convert the signal photon to the common output at $\omout$. The output mode is strongly coupled to the bus waveguide causing the converted photon to exit the resonator and couple into the output waveguide through the MZI filter.

We stress that the switches only operate on the classical pump fields $-$ making their insertion loss far less critical than if they operated on single photons.

\subsection{Interferometrically Coupled Resonator\seclab{FWM resonator}} 
The FWM-resonator is sketched again in~\figref{props}b with the fields, $s$, in various parts of the device indicated. Considering a case without input fields ($\ssub{s}{\rm{in}}\equal 0$) and a field, $s_g$, being generated inside ring 1, the intra-cavity field of ring 1 is (see~\appref{FDM mzi} for details)
\begin{align}\eqlab{intra cavity field IC}
\frac{\ssub{s}{1-}}{\ssub{s}{g}} =  \frac{1}{1 - \vsubsup{\overline{C}}{1,1}{(\mathcal{I},1)}e^{i\Phi_1}\ssub{\overline{t}}{12} } , ~~~~~ \ssub{\bar{t}}{12} =  \frac{\ssub{\nu}{2} - e^{i\ssub{\Phi}{2}}} {1 - \ssub{\nu}{2}  e^{i\ssub{\Phi}{2}} },
\end{align}
where $\Phi_1$ ($\Phi_2$) is the round-trip phase of ring 1 (2). The through-coupling coefficient of the coupling region between the rings is $\ssub{\nu}{2}$. The matrix describing the MZI coupling region is
\begin{align}\eqlab{coupling matrix mzi}
\vsup{\overline{C}}{(\mathcal{I},1)} = \left[\!\!\begin{array}{c c} \ssub{\nu}{1}^2 - e^{i\psi}( 1 - \ssub{\nu}{1}^2) 	& i\ssub{\nu}{1}\sqrt{1\minus\ssub{\nu}{1}^2} \Big(1\plus e^{i\psi}\Big) \\
i\ssub{\nu}{1}\sqrt{1\minus\ssub{\nu}{1}^2} \Big(1\plus e^{i\psi}\Big)  	& \ssub{\nu}{1}^2\Big(1\plus e^{i\psi}\Big) - 1  \end{array} \!\!\right],
\end{align}
where $\psi$ is the phase imbalance of the MZI and the directional couplers are assumed identical with a through-coupling coefficient $\ssub{\nu}{1}$. 
The blue curve in~\figref{props}c plots~\eqref{intra cavity field IC} and the dashed red curve plots the off-diagonal elements of the MZI coupling matrix in~\eqref{coupling matrix mzi} for a FWM resonator design where the length of ring 1 is four times larger than ring 2 and six times larger than the path-length difference of the coupling-interferometer. 
%
\begin{figure}[!h]
  \centering
  \includegraphics[height=6cm]{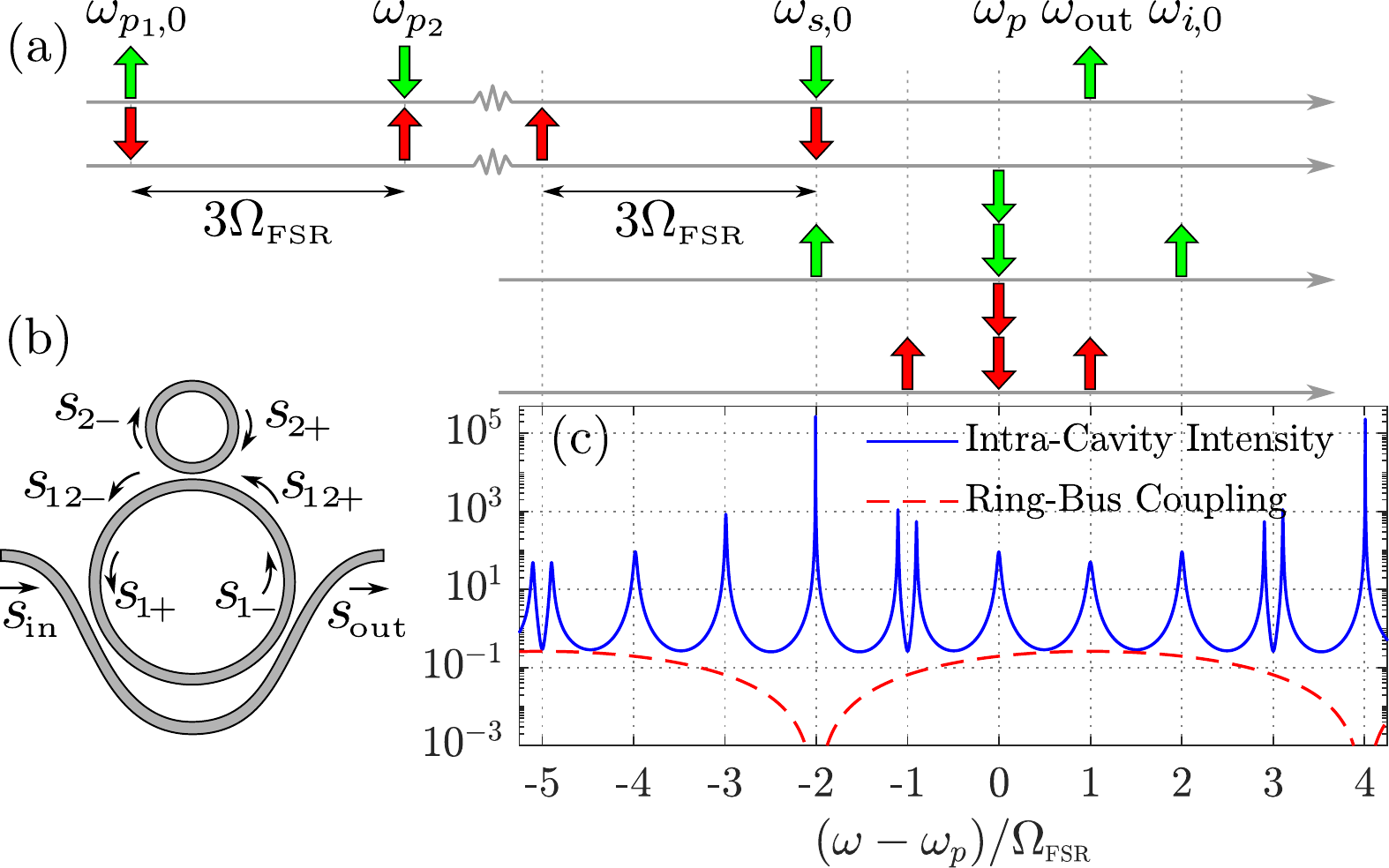}
 \caption{(a) Illustration of FWM processes occuring in the resonator shown in (b). Arrows pointing up correspond to creation of a photon whereas downwards-pointing arrows correspond to annihilation of a photon. Suppressed processes involving modes with resonance-splitting are indicated by red arrows and desired processes are indicated by green arrows. (b) Sketch of the FWM-resonator. (c) Intra-cavity intensity of ring 1 in~\eqref{intra cavity field IC} and ring-bus coupling in~\eqref{coupling matrix mzi}. Note that (a) and (c) share the frequency axis and BS-FWM pumps are far-detuned from other modes as in Ref.~\cite{Li2016}.}
\figlab{props}
\end{figure}
Resonances of ring 1 are enumerated relative to the SFWM pump as $\omSub{j}\equal \omSub{p}\plus j\OmFSR$ with $j\!\in\!\mathbb{Z}$ and ring 2 has resonances at $\omSub{\minus 1\plus 4n}$ with $n\!\in\!\mathbb{Z}$.~\figref{props}c shows that signal modes at $\omSub{s,n} \equal \omSub{-2\plus 6n}$ are decoupled from the bus waveguide whereas the pump ($\omSub{p}$), idler ($\omSub{i,n} \equal \omSub{2-6n}$), and output mode ($\omout\equal \omSub{1}$) are strongly coupled. Note that $n$ attains both positive and negative values so that signal modes exist on both sides of $\omSub{p}$.~\figref{props}a illustrates the relevant FWM processes of the pair creation and frequency conversion for $n\equal 0$. Direct generation of photons at $\omout$ $(2\omSub{p} \rightarrow \omout + \omSub{-1})$ must be avoided since emission should only occur after step two of the protocol. The process is indeed suppressed due to mode-splitting at $\omSub{-1}$. Additionally, mode-splittings at $\omSub{\minus 5\plus 12n}$  ensure suppression of the conversion process $\omSub{s,n}+\omSub{p_1,n} \!\rightarrow  \omSub{\minus 5\plus 12n} + \omSub{p_2}$, whereas the desired multiplexing conversion in the opposite direction, $\omSub{s,n}+\omSub{p_2} \!\rightarrow \omout + \omSub{p_1,n}$, is resonantly enhanced. 

In the high conversion efficiency regime, it becomes important to consider cascaded BS-FWM processes in which photons get converted several times before coupling into the bus waveguide. This severely limits the conversion efficiency without interferometric coupling. In~\appref{cascaded FWM} we show that the conversion efficiency is limited to 50\% even for perfect unidirectionality by including first order cascaded FWM. However, we note that signal modes in~\figref{props}c are located symmetrically around $\omout$, such that 
\begin{align}\eqlab{cascade suppression}
\big|\omSub{s,n}-\omout\big| = \big|\omout - \omSub{s,-(n-1)}\big|, ~~n\!\in\!\mathbb{Z}.
\end{align}
This ensures that the first-order cascaded BS-FWM process is $\omSub{s,n}\rightarrow \omout \rightarrow \omSub{s,-(n-1)}$, where the photon ends on another signal mode. Any higher order process is suppressed by mode-splitting and the photon is eventually converted back to $\omout$. 

Undesired SFWM producing photons at $\omout$ or $\omSub{s,n}$ from the BS-FWM pumps may be suppressed by placing the BS-FWM and SFWM pumps on either side of the zero-dispersion wavelength matching their group indices~\cite{Li2016}. In this way, the undesired SFWM processes are not phase-matched and therefore suppressed. If the path-length difference of the MZI-filter in~\figref{PIC} is two times shorter than ring 1 it may be adjusted to separate odd- from even-numbered modes such that $\omout$ is dropped to the output waveguide while $\omSub{p}$ and all idler modes continue in the bus waveguide towards the SNSPDs.

\subsection{Spectral Correlations of Generated Photon Pairs\seclab{pair production}}
An important advantage of our proposed design is its ability to produce photons with very high spectral purity. It has been shown that the spectral purity of photons emitted from a resonator where idler, pump, and signal modes are identical is limited to 92\% even for a flat pump spectrum~\cite{Helt2010, Vernon2017}. In Ref.~\cite{Vernon2017} it was shown that using interferometric coupling to increase the linewidth of the pump relative to the signal and idler enables arbitrarily high spectral purity. Another solution is to modify the pump spectrum as shown in Ref.~\cite{Christensen2018}. Our design achieves a signal linewidth, $\gamSub{s}$, that is smaller than the pump, $\gamSub{p}$, and idler, $\gamSub{i}$.
~\figref{state_purity}a plots the spectral purity of our device as a function of the ratio between the quality factors ($Q_j\equal \omega_j/\gamSub{j}$ with $j$ labeling the resonance) of the signal and idler modes when assuming $Q_p\equal Q_i$ and a flat pump spectrum (see~\appref{pair production app} for details).
\begin{figure}[!h]
  \centering
  \includegraphics[height=3.7cm]{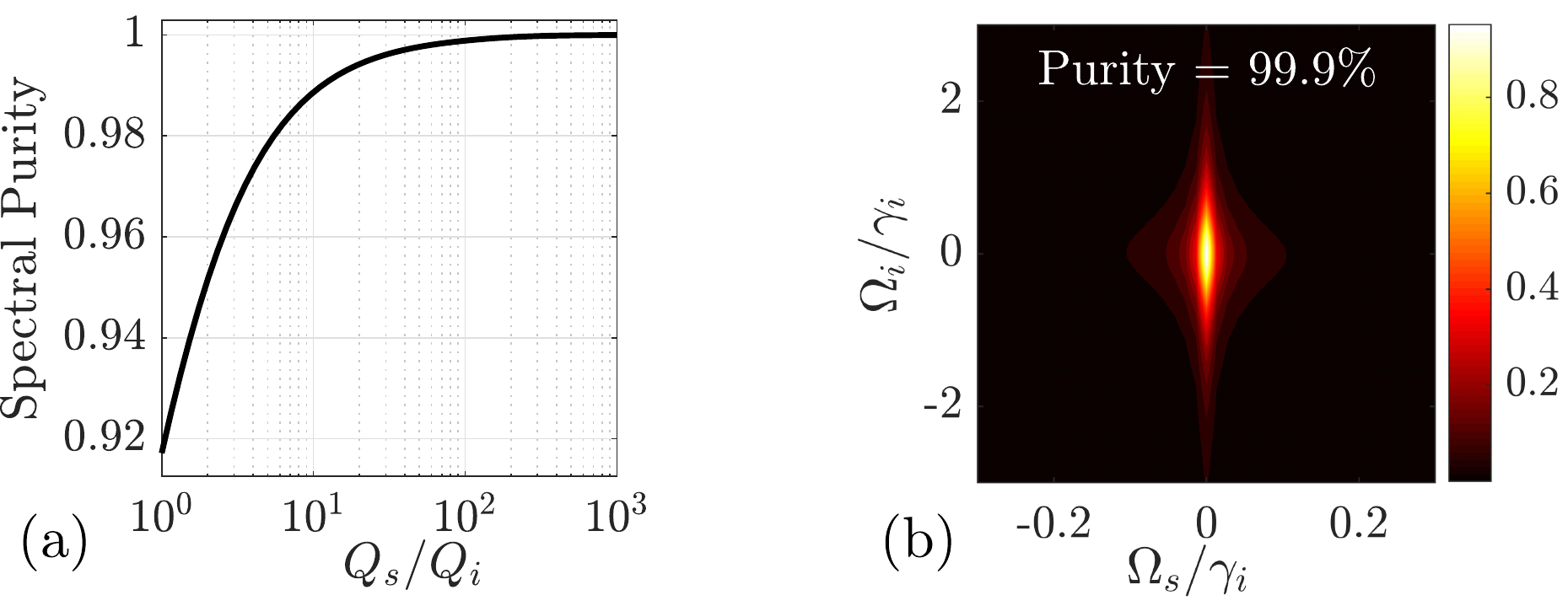}
 \caption{(a) The maximum heralded purity versus the ratio of signal and idler quality factors. (b) The joint spectral intensity for $Q_s/Q_i = 100$. In both (a) and (b) it was assumed that $Q_p\equal Q_i$ and that the pump spectrum was flat. Note that we assumed zero intrinsic loss in these calculations.}
\figlab{state_purity}
\end{figure}
We note that this is the purity of signal photons being leaked into the environment and leave the purity analysis of photons coupled out after frequency conversion for future work.~\figref{state_purity}a shows that the spectral purity rapidly increases towards unity as $Q_s/Q_i$ increases. In the next section, we show that the frequency conversion efficiency increases with $Q_L/Q_i$ (were $Q_L\equal \omega_p/\gamSub{L}$), which means that the spectral purity increases simultaneously since $Q_s\approx Q_L$ for the interferometrically coupled device. For $Q_s/Q_i = 100$, the resulting joint spectral intensity is plotted in \figref{state_purity}b. It displays a broad idler distribution and narrow signal distribution, with negligible correlations between signal and idler frequencies, giving rise to a spectral purity of $99.9 \%$.

\subsection{Photon Frequency Conversion\seclab{frequency conversion}}
The second step of the emission protocol consists of frequency converting a signal photon from mode ($s$) to the output mode (denoted $o$ here for brevity). For a material without TPA, the efficiency is only limited by the ring-ring coupling in the form $G\equal g/\sqrt{\gamSub{L}(\gamSub{L}\plus\gamSub{o})}$ and the ratio between the coupling- and loss rates, $Q_L/Q_o$. Here, we are interested in emitting photons into a specific wave packet described by the function $\So{o}(t)$. Photons with time-symmetric wave packets are particularly interesting for e.g. two-photon gates~\cite{Nysteen2017} and reducing sensitivity to timing jitter in two-photon interference~\cite{Branczyk2017} so we consider a Gaussian waveform as a specific example 
\begin{align}\eqlab{Gaussian main}
\So{o}^{\rm{Gauss}}(t) =  \sqrt{\frac{2}{\Delta t}} \left(\frac{\text{ln}(2)}{\pi}\right)^{\!\frac{1}{4}} \exp\!\left(\!-2\text{ln}(2)\frac{(t-t_0)^2}{\Delta t^2} \right).
\end{align} 
It has a temporal full width at half maximum (FWHM) $\Delta t$ and spectral width $\Delta\omega\equal 4\text{ln}(2)/\Delta t$. 
\begin{figure}[!h]
  \centering
  \includegraphics[height=3.3cm]{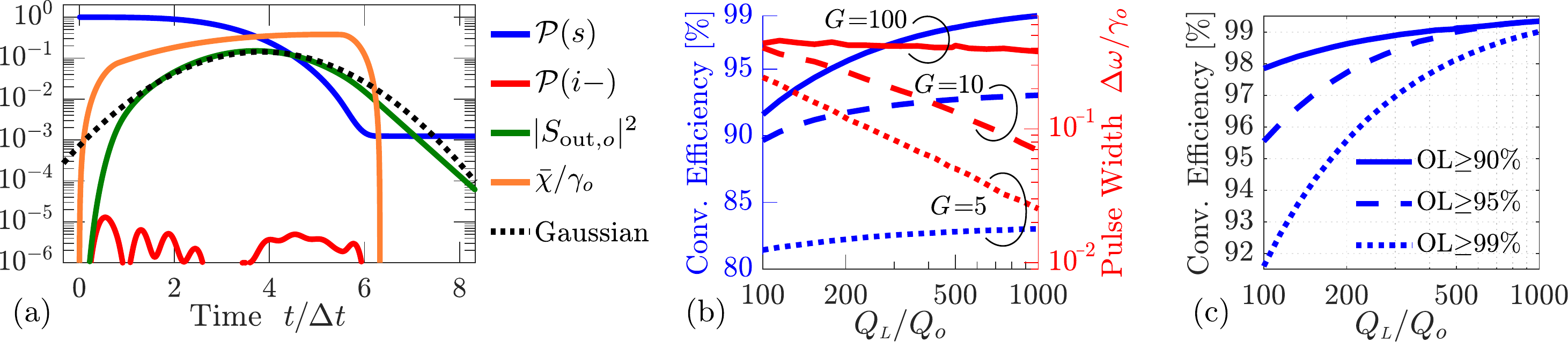}
 \caption{(a) A solution of the emitted wave packet (green) along with the occupation probability of the signal mode (blue) and down-converted mode (red), as well as $\bar{\chi}(t)$ (orange). (b) Conversion efficiency (blue, left axis) as a function of $Q_L/Q_o$ for different values of the ring-ring coupling, $G$. The corresponding optimum pulse widths are plotted in red (right axis). These values were obtained using the criteria $\text{OL}\geq 99\%$. (c) Conversion efficiency as a function of $Q_L/Q_o$ for $G\equal 100$ for different overlap requirements. }
\figlab{Gaussian_Emission props}
\end{figure}
In~\appref{Shaping Output Photons}, we derive the temporal shape of $\bar{\chi}(t)$ required to achieve the desired output wave packet.~\figref{Gaussian_Emission props}a plots the solution for a specific case of $Q_L/Q_o\equal 500$, $G\equal 100$, and  $\Delta\omega/\gamSub{o}\equal 0.38$. The deviation of the output, $\So{o}$, from the desired wave packet, $\So{o}^{\rm{Gauss}}$, is quantified by the conversion efficiency
\begin{align}\eqlab{conversion efficiency emission main}
	\eta_{\text{out}} &= \int_0^{\infty} \!|\So{o}(t)|^2 dt ,
\end{align} 
and overlap 
\begin{align}\eqlab{overlap definition main}
	\text{OL} &= \langle \So{o} | \So{o}^{\rm{Gauss}} \rangle = \frac{1}{\eta_{\text{out}}} \left|\int_0^\infty \So{o}^*(t) \So{o}^{\rm{Gauss}}(t) dt  \right|^2 .
\end{align} 
To investigate the influence of loss and out-coupling through the down-converted mode, we evaluate the figures of merit for different values of $Q_L/Q_o$ and $G$. The results are shown in Figs. \ref{fig:Gaussian_Emission props}b,c. In~\figref{Gaussian_Emission props}b we require an overlap of at least 99\%, and it is seen that 99\% conversion efficiency is achievable with $G\equal 100$ and $Q_L/Q_o\sim \!1000$. As $G$ decreases, the bandwidth of large extinction between up- and down-conversion also decreases. This necessitates narrow bandwidth pulses, which are longer in time leading to higher loss during emission. Therefore, both the conversion efficiency and optimum pulse bandwidth decreases with decreasing $G$ as illustrated in~\figref{Gaussian_Emission props}b.

If the overlap requirement is relaxed, the conversion efficiency can be significantly increased as illustrated in~\figref{Gaussian_Emission props}c. This demonstrates that there exists a trade-off between conversion efficiency and the desired temporal wave packets of the emitted photons, which is important to keep in mind for different applications of the source.

\subsection{ Adding Temporal Multiplexing\label{sec:temporal multiplexing}}
A total of $N$ spectral multiplexing modes requires $N$ drop-filters and SNSPDs as well as $N\!+\!2$ SOI switches and external lasers. The associated fabrication complexity of adding modes is minimal as it simply corresponds to extending the rows of SNSPDs and switches in~\figref{PIC}. We note that temporal multiplexing-modes may be introduced in parallel by repeating step one of the protocol $M$ times in each emission cycle. In this case, the control signals from the SNSPDs also need to flip the switch controlling the passage of the SFWM pump to avoid subsequent pair generation. Switches based on free carrier dispersion~\cite{Gehl2017} with a carrier relaxation time significantly longer than the duration of one emission cycle would contain the necessary memory effect to include temporal modes without requiring a memory in the superconducting circuit. Such two-dimensional multiplexing opens up trade-offs between resource requirements and system properties since temporal modes do not require additional resources but losses increase exponentially with $M$. It is therefore possible to optimize $N$ and $M$ for a given set of system parameters and fabrication yield to maximize the probability of successful state preparation. However, the ideal scenario would be to use as many spectral modes as possible since temporal modes introduce loss and effectively reduces the emission rate.

The number of available spectral multiplexing modes depends on the bandwidth over which $|\omSub{p_2} -\omSub{p_1,n}| \approx |\omout - \omSub{s,n}|$, relative to the FSR of ring 1. $N\equal 16$ is the lower limit to achieve 99\% fidelity of the single-photon state assuming zero loss and number-resolving detectors~\cite{Christ2012}. It seems feasible that 10-100 multiplexing modes is possible in SiN microrings~\cite{Okawachi2011}. 

\section{ Discussion}\label{sec:discussion}
With our experimental demonstration we have shown that unity frequency conversion efficiency is feasible among the dense frequency modes of PIC ring resonators. Achieving on-chip multiplexing requires a latency of the quantum feedback below the photon storage time, which is only possible by placing switches in close proximity to the SNSPDs. The SOI-based switches must then be functional at cryogenic temperatures, which has recently been demonstrated~\cite{Gehl2017}. Driving the switches using the weak electrical signals output from the SNSPDs is challenging, but devices capable of amplification and logic operations have been developed~\cite{McCaughan2014}. Single chip filtering with sufficient extinction for pump rejection has recently been demonstrated~\cite{Gentry2018}. However, the BS-FWM pumps must be even higher power than the SFWM pump and further progress in on-chip filtering is necessary. Additionally, it is necessary to statically tune the rings of the FWM resonator as well as the drop filters. Opto-electro-mechanical tuning of the refractive index in waveguides is suitable for cryogenic operation and demonstrations of large index shifts with low loss have been presented~\cite{Pruessner2016}. \\

In conclusion, we have experimentally demonstrated that unidirectional frequency conversion between modes of ring resonators is possible with more than 40$\,$dB extinction. Our theoretical investigation shows that this leads to near-unity conversion efficiency and based on this, we proposed a scheme for PIC frequency multiplexed single photon sources with high performance. For instance, Figs.~\ref{fig:state_purity}b and~\ref{fig:Gaussian_Emission props}c show that 99\% conversion efficiency and spectral purity is possible for $Q_L/Q_i\!\sim \!400$. If, for instance, the pump, idler, and output modes have coupling $Q$s of $10^4$ the corresponding intrinsic $Q$ must be $ 4\!\times\!10^6$, which is well below what has been demonstrated~\cite{Ji2017}. For an analysis of the full system efficiency including latency in the quantum feedback as well as detector efficiency, we refer to Ref.~\cite{Heuck2018}. Importantly, we note that extraction of the created signal photon from the ring is included in the frequency conversion efficiency here, whereas the loss associated with this process was not treated in Ref.~\cite{Heuck2018}. We consider our proposal to be a very promising route to on-chip multiplexed single-photon sources for near-term implementation. The constituent components have been demonstrated individually~\cite{Shainline2017,Gehl2017,McCaughan2014,Gentry2018, Pruessner2016} and switching only the classical fields significantly improves the loss-budget over other types of multiplexing. 


\section*{Acknowledgements}
This work was supported by the Danish National Research Foundation through the Center of Excellence SPOC (Silicon Photonics for Optical Communication), DNRF 123. M. H. acknowledges funding from VILLUM FONDEN.

\setcounter{equation}{0}
\setcounter{figure}{0}
\setcounter{table}{0}
\setcounter{section}{0}
\renewcommand{\theequation}{A\arabic{equation}}
\renewcommand{\thefigure}{A\arabic{figure}}
\renewcommand{\thetable}{A\arabic{table}}
\renewcommand{\thesection}{Appendix \Alph{section}}


\section{Device Models}\label{app:unidirectional theory}
It is useful to consider two descriptions of the coupled resonators in this work. One, which we denote the coupled-mode model (CMM), is convenient for modeling the dynamics of the system and predict its performance. The other, denoted the frequency domain model (FDM), is useful for device design.~\figref{two ring}a shows a sketch of the device indicating parameters in the CMM and~\figref{two ring}b shows the fields used in the FDM analysis.
\begin{figure}[!h]
  \centering
  \includegraphics[height=3.0cm]{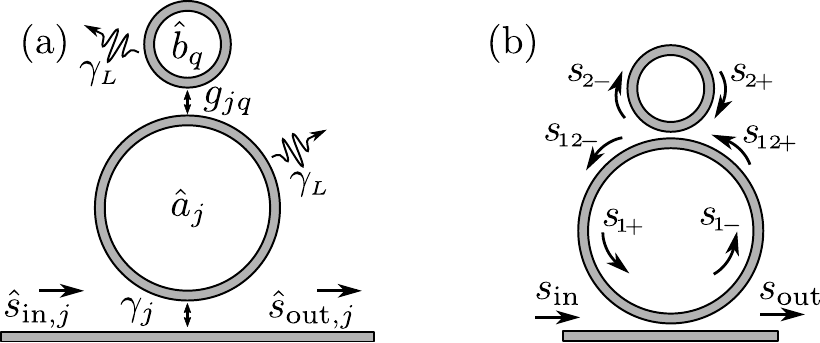}
 \caption{(a) Sketch of the device illustrating parameters of the coupled-mode model. (b) Sketch of the device showing fields used in the frequency domain model.}
\figlab{two ring}
\end{figure}
Below, we go through each model description and explain how to relate their parameters. This ensures that geometrical properties of devices can be related to their performance and thereby assist the design process.

\subsection{Coupled-Mode Model}\label{app:CMM}
The CMM is the standard description used in open quantum systems where the cavity resonances are treated as discrete modes that couple to each other and the continuous waveguide-modes with coupling rates $g_{jq}$ and $\gamSub{j}$, respectively. The system is mode\-led by the Hamiltonian $H\equal H_L\plus H_{NL}$ with~\cite{Combes2017}
\begin{multline}\eqlab{Hamiltonian linear}
H_L 	=  \sum_j \ssub{\omega}{j}\hatdsub{a}{j}\hatsub{a}{j} + \sum_j i\sqrt{\frac{\ssub{\gamma}{j}}{2\pi}}\int_{-\infty}^{\infty}\!\!\!d\omega \big[\hatsub{a}{j}\hatdsub{s}{}(\omega) - \hatdsub{a}{j}\hatsub{s}{}(\omega)\big]  ~+\\
\int_{-\infty}^{\infty} \!\!\!d\omega \omega \hatdsub{s}{}(\omega)\hatsub{s}{}(\omega) + \ssub{\omega}{b}\hatdsub{b}{}\hatsub{b}{} + g\big(\hatdsub{a}{i-}\hatsub{b}{} + \hatdsub{b}{}\hatsub{a}{i-}\big),
\end{multline} 
where $j\in \{s,i-,i+\}$. The nonlinear part is
\begin{align}\eqlab{Hamiltonian nonlinear}
H_{NL} 	= \ssub{\chi}{} \big( \hatdsub{a}{1}\hatdsub{a}{i+}\hatsub{a}{2}\hatsub{a}{s} +   \hatdsub{a}{2}\hatdsub{a}{i-}\hatsub{a}{1}\hatsub{a}{s} \big) + h.c.
\end{align} 
The heat bath responsible for the loss rate, $\gamSub{L}$, is not explicitly included in $H_L$ but the loss rates are included in operator equations of motion (see Ref.~\cite{Vernon2015a} for the necessary derivations).
As illustrated in~\figref{two ring}a, the modes in ring 1, ring 2, and the bus waveguide are represented by the annihilation operators $\hatsub{a}{j}$, $\hatsub{b}{q}$, and $\hatsub{s}{j}$, respectively.  We only consider one mode of ring 2, which only couples to mode $i-$ of ring 1, so we have dropped the subscripts on $\hatsub{b}{}$ and $g$ in~\eqref{Hamiltonian linear}.

Equations of motion for the electric fields of the cavity modes may be found using the Hamiltonian in~\eqsref{Hamiltonian linear}{Hamiltonian nonlinear}~\cite{Vernon2016}
\begin{subequations}\eqlab{EOM 1}
\begin{align}
	\ssub{\dot{A}}{s} 	&= - \frac{\GamSub{s}}{2} \ssub{A}{s} -i\bar{\chi}^{*}\ssub{A}{i+} -i\bar{\chi}\ssub{A}{i-} - \sqrt{\gamSub{s}}\Si{s} \eqlab{EOM as 1}\\
	\ssub{\dot{A}}{i+} 	&= - \frac{\GamSub{i+}}{2}\ssub{A}{i+} - i\bar{\chi} \ssub{A}{s} - \sqrt{\gamSub{i+}}\Si{i+}  \eqlab{EOM ai+ 1}\\	
    \ssub{\dot{A}}{i-}	&=  - \frac{\GamSub{i-}}{2}\ssub{A}{i-} - i\bar{\chi}^{*}\ssub{A}{s} -ig\ssub{B}{} - \sqrt{\gamSub{i-}}\Si{i-} \eqlab{EOM ai- 1} \\ 
    \ssub{\dot{B}}{}	&= \left(-i\delSub{ab} - \frac{\gamSub{L}}{2} \right)\ssub{B}{} -ig\ssub{A}{i-} \eqlab{EOM b 1}\\
    \So{j} &= \Si{j} + \sqrt{\gamSub{j}}\ssub{A}{j}, \quad j \!\in\! \{s, i-, i+\}. \eqlab{EOM IO 1} 
\end{align} 
\end{subequations}
The fields in~\eqref{EOM 1} are slowly varying amplitudes defined with reference to the mode resonances $\ssub{a}{j}(t)\equal\ssub{A}{j}(t)\exp(-i\omSub{j}t)$ and $s_{\text{in},j/\text{out},j}(t)\equal S_{\text{in},j/\text{out},j}(t)\exp(-i\omSub{j}t)$. The field in ring 2 is $\ssub{b}{}(t)\equal\ssub{B}{}(t)\exp(-i\omSub{i-}t)$, which gives rise to the detuning $\delSub{ab}\equal \omSub{b}-\omSub{i-}$ in~\eqref{EOM b 1}. The time-dependent nonlinearity due to the BS-FWM pumps is $\bar{\chi}^{*}\equal \chi \braket{\hatsub{a}{p_1}\hatdsub{a}{p_2}}$ and $\GamSub{j} \equal \gamSub{j}+\gamSub{L}$ with $j\!\in\! \{s,i-,i+\}$. In the linear regime, $\bar{\chi}\equal 0$,~\eqsref{EOM ai- 1}{EOM b 1} decouple from the rest and describe two coupled resonators. In the strong coupling regime, $g\!>\!\gamSub{i-}/4$, linear superpositions of $\ssub{A}{i-}$ and $\ssub{B}{}$ form uncoupled super-modes with modified re\-sonance frequencies. If $\omSub{i-}\equal \omSub{b}$, the eigenfrequencies of the super-modes are shifted by $\pm\sqrt{g^2-(\gamSub{i-}/4)^2}$ relative to the degenerate resonances of $\ssub{A}{i-}$ and $\ssub{B}{}$. This mode-splitting is observed in Fig. 2b for every fourth mode of ring 1 because the resonances of each ring are aligned and the free-spectral-range (FSR) of ring 2 is four times larger than that of ring 1. 

To analyze the frequency conversion properties of the device, we make the simplifying assumptions $\gamSub{i+}\equal\gamSub{i-}\equal\gamSub{i}$  and $\bar{\chi} \equal \bar{\chi}^{*}$. For continuous wave (CW) pump fields,~\eqref{EOM 1} may be solved using Fourier transforms. From the solution, we define the extinction ratio
\begin{align}\eqlab{isolation ratio 1}
\zeta(\Omega) = \frac{\Sto{i+}(\Omega)}{\Sto{i-}(\Omega)}  = \frac{4g^2 }{\big[\gamSub{L}+i2(\delSub{ab}-\Omega)\big]\big(\GamSub{i}-i2\Omega\big)} + 1,
\end{align}
where $\Omega$ is the frequency separation of each mode from its resonance, $\ssub{\tilde{A}}{j}(\Omega)\equal \ssub{\tilde{a}}{j}(\Omega \plus \omSub{j})$, as well as the separation of the input field from the signal resonance, $\Sti{s}(\Omega)\equal \sti{s}(\Omega\plus\omSub{s})$. The $\tilde{}$ is used to indicate frequency domain fields. The maximum extinction is found for $\Omega\equal 0$ and $\delSub{ab}\equal 0$
\begin{align}\eqlab{isolation ratio max}
\zeta^{\rm{max}} = \frac{4g^2 }{\gamSub{L}\GamSub{i}} + 1 = 4G^2+1,
\end{align}
where the normalized coupling parameter $G\equiv g/\sqrt{\gamSub{L}\GamSub{i}}$ was defined. The up-conversion efficiency, 
\begin{align}\eqlab{eta CW app}
	\eta_{i+} &\!=\!  \left|\frac{\Sto{i+}}{\Sti{s}}\right|^2 \!\!\!=\! \left| \frac{4\sqrt{\gamSub{i}\gamSub{s}}\bar{\chi}}{(\GamSub{i} \minus i2\Omega)(\GamSub{s} \minus i2\Omega) \plus 4\bar{\chi}^2\big(1 \plus \zeta^{-1}\big)} \right|^2 ,
\end{align} 
has a maximum of
\begin{align}\eqlab{eta CW max app}
\eta_{i+}^{\rm{max}} =  \frac{1 + 4G^2}{2 + 4G^2} \frac{\gamSub{i}\gamSub{s}}{\GamSub{i}\GamSub{s}} ,
\end{align} 
which is achieved when $\delSub{ab}\equal \Omega\equal 0$ and for a nonlinearity
\begin{align}\eqlab{chi max}
\bar{\chi}^{\rm{max}} =  \frac12 \sqrt{\frac{1 + 4G^2}{2 + 4G^2}} \sqrt{\GamSub{i}\GamSub{s}}.
\end{align} 
\eqsref{eta CW app}{eta CW max app} are identical to Eqs. (28) and (32) in Ref.~\cite{Vernon2016} in the limit $G\!\rightarrow\!\infty$ (note the factor of 2 difference in our definition of decay rates). In the limit of large $G$ and highly over-coupled ring modes,~\eqref{eta CW max app} may be written as
\begin{align}\eqlab{eta CW max app scaling}
\eta_{i+}^{\rm{max}} \approx  \left(1 - \frac{1}{4G^2}\right) \left( 1 - \frac{Q_s+Q_i}{Q_L} \right) ,
\end{align} 
which clearly shows the scaling with $G$ and the ratio of coupling- to intrinsic quality factor ($Q_j\equal \omSub{j}/\gamSub{j}$). Reaching 99\% conversion efficiency requires $G\sim \!10$ and $Q_L/Q_j\sim \!100$ with $j\!\in\!\{s,i\}$. The linear transmission in the vicinity of $\omSub{i-}$ and $\omSub{i+}$ is
\begin{subequations}\eqlab{trans CW CMT 1}
\begin{align}\eqlab{trans CW CMT i- 1}
t_{i-}(\Omega)  &= \frac{\So{i-}(\Omega)}{\Si{i-}(\Omega)} = 1-\frac{\gamSub{i-}}{ \frac{\GamSub{i-}}{2} - i\Omega + \frac{g^2}{\gamSub{L}/2 - i(\Omega-\delta_{ab}) }} \\
t_{i+}(\Omega)  &= \frac{\So{i+}(\Omega)}{\Si{i+}(\Omega)} = 1-\frac{\gamSub{i+}}{    \frac{\GamSub{i+}}{2} - i\Omega }, \eqlab{trans CW CMT i+ 1}
\end{align}
\end{subequations}
where $\Omega$ again is the frequency separation from $\omSub{i-}$ in~\eqref{trans CW CMT i- 1} and from $\omSub{i+}$ in~\eqref{trans CW CMT i+ 1}. The transmission spectra in~\eqref{trans CW CMT 1} are found by assuming all input fields in~\eqref{EOM 1} are zero except $\Si{i-}$ for~\eqref{trans CW CMT i- 1} and $\Si{i+}$ for ~\eqref{trans CW CMT i+ 1}.

\subsection{\label{app:FDM} Frequency Domain Model}
In the frequency domain model, the fields everywhere in the device (see~\figref{two ring}b) are connected by transfer matrices. We assume the directional couplers are well described by a matrix, $\vsup{C}{(n)}$, such that
\begin{subequations}\eqlab{field relations}
\begin{align}
\left[\!\!\!\begin{array}{c} \ssub{s}{1-}  	\\
\ssub{s}{\rm{out}} \end{array} \!\!\!\right] &= \vsup{C}{(1)} \left[\!\!\!\begin{array}{c} \ssub{s}{1+}  	\\
\ssub{s}{\rm{in}} \end{array} \!\!\!\right] ,\quad
		\left[\!\!\!\begin{array}{c} \ssub{s}{2-}  	\\
\ssub{s}{12-} \end{array} \!\!\!\right] = \vsup{C}{(2)}  \left[\!\!\!\begin{array}{c} \ssub{s}{2+}  	\\
\ssub{s}{12+} \end{array} \!\!\!\right] \\
	\ssub{s}{12+} &= \ssub{s}{1-}e^{i\frac{\phi_1}{2}}, ~~ \ssub{s}{1+} = \ssub{s}{12-}e^{i\frac{\phi_1}{2}}, ~~ \ssub{s}{2+} = \ssub{s}{2-}e^{i\phi_2} .
\end{align}
\end{subequations}
Solving~\eqref{field relations} we find the transmission, $\ssub{t}{12}$, of the intra-cavity field in ring 1 when passing ring 2 and the transmission, $t$, through the bus waveguide 
\begin{align}
\ssub{t}{12} &= \frac{\ssub{s}{12-}}{\ssub{s}{12+}} =  \vsubsup{C}{2,2}{(2)} + \frac{\vsubsup{C}{1,2}{(2)} \vsubsup{C}{2,1}{(2)}e^{i\phi_2}}{1 - \vsubsup{C}{1,1}{(2)}e^{i\phi_2} } \eqlab{trans t12}\\
t &= \frac{\ssub{s}{\rm{out}}}{\ssub{s}{\rm{in}}} =  \vsubsup{C}{2,2}{(1)} + \frac{\vsubsup{C}{1,2}{(1)} \vsubsup{C}{2,1}{(1)}e^{i\phi_1} \ssub{t}{12}}{1 - \vsubsup{C}{1,1}{(1)}e^{i\phi_1}\ssub{t}{12} } . \eqlab{trans t}
\end{align}
The coupling regions are assumed to be described by symmetric transfer matrices
\begin{align}\eqlab{coupling matrix}
\vsup{C}{(n)} = e^{i\ssub{\theta}{n}} \left[\!\begin{array}{c c} \ssub{\nu}{n}  	& i\sqrt{1-\ssub{\nu}{n}^2} \\
i\sqrt{1-\ssub{\nu}{n} ^2} 	& \ssub{\nu}{n}   \end{array} \!\right],
\end{align}
where $\ssub{\theta}{n}$ is the phase accumulated along the directional coupler and $\nu_n$ is the through-coupling coefficient. The phase accumulated through a length $L$ is $k(\omega)L$ with the propagation constant approximated by
\begin{align} \eqlab{k approx}
k(\omega) \approx k(\omref) + \frac{\partial k}{\partial \omega}\Delta\omega = \frac{ \tilde{n}_{\rm{eff}} \omref }{c} +  \frac{n_g }{c}\Delta\omega,
\end{align}
where $\Delta\omega\equal \omega-\omref$. The imaginary part of the complex refractive index, $\tilde{n}_{\rm{eff}}\equal n_{\rm{eff}}' + i n_{\rm{eff}}''$, is related to the linear amplitude loss coefficient and intensity loss rate by
\begin{align}  \eqlab{complex ref index}
\ssub{\alpha}{L} = \frac{\omref n_{\rm{eff}}''}{c} = \frac{\ssub{\gamma}{L}}{2} \frac{n_g}{c} .
\end{align}
The round-trip loss of the circulating field in ring 1 due to coupling to the bus waveguide is $|\vsubsup{C}{1,1}{(1)}|\equal \nu_1$ and a connection to the CMM is given by
\begin{align}  \eqlab{waveguide coupling relation}
\exp[-\gamSub{1}\tau_{\rm{\scriptscriptstyle RT,1}}] = \nu_1^2~~\Rightarrow ~~\gamSub{1}\equal -\frac{c}{n_gL_1}\ln\!\big(\nu_1^2\big),
\end{align}
where $\tau_{\rm{\scriptscriptstyle RT,1}}$ is the round-trip time of ring 1. The coupling rate between the rings, $g$, can be related to the parameters of the FDM by considering the modification that ring 2 imposes on the circulating field in ring 1. Every time the field passes by ring 2 it acquires the amplitude and phase contained in $t_{12}$ of~\eqref{trans t12}. This is seen from the round-trip term, $\vsubsup{C}{1,1}{(1)}e^{i\phi_1}\ssub{t}{12}$, in the denominator on the right hand side of~\eqref{trans t}. Defining $\vsup{\overline{C}}{(n)} \!\equal \vsup{C}{(n)}e^{-i\theta_n}$, and $\ssub{\overline{t}}{12} \equal \ssub{t}{12}e^{-i\theta_2}$, the round-trip term is $\vsubsup{\overline{C}}{1,1}{(1)}e^{i\Phi_1}\ssub{\overline{t}}{12}$, where $\Phi_1 \equal \phi_1 \plus \theta_1 \plus \theta_2$ is the round-trip phase of ring 1 without coupling between the rings. In this form, it is clear that the phase of $\ssub{\overline{t}}{12}$ modifies the resonance condition of ring 1.~\figref{model comparison}a plots the phases of the round-trip term when both rings have a resonance at $\omSub{i-}$.
\begin{figure}[!h]
  \centering
  \includegraphics[height=4cm]{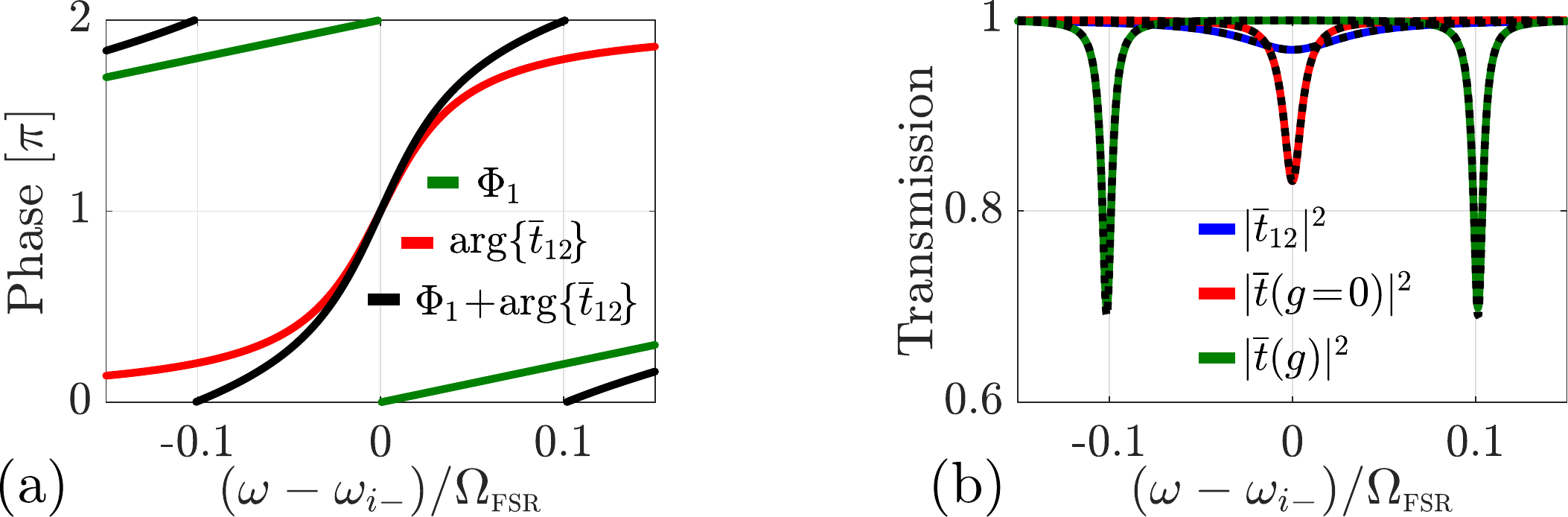}
 \caption{(a) Phases of the round-trip factor, $\vsubsup{\overline{C}}{1,1}{(1)}e^{i\Phi_1}\ssub{\overline{t}}{12}$, as a function of frequency. (b) Comparison of the transmission calculated from~\eqsref{trans t12}{trans t} (blue, red, green) and~\eqsref{trans CW CMT i- 1}{trans CW CMT i+ 1} (dotted black). The used parameters are: $G\equal 16.9$, $\gamSub{1}\equal 0.011\OmFSR$, $\gamSub{2}\equal 0.067\OmFSR$, and $\gamSub{L}\equal 5.3\!\times\! 10^{-4}\OmFSR$.}
\figlab{model comparison}
\end{figure}
Since $\arg\{\ssub{\overline{t}}{12}(\omSub{i-}\!)\}\equal \pi$, the coupled system is anti-resonant at $\omSub{i-}$ and the black curve shows that two new resonances appear at $\omSub{i-}\pm 0.1\OmFSR$ where the total round-trip phase equals $2\pi$. The FSR of ring 1 is $\OmFSR\equal 2\pi c/(n_g L_1)$. The FDM therefore offers an interpretation of the mode-splitting in terms of dispersion engineering, which was also employed in Refs.~\cite{Gentry2014,Xue2015}. The parameters $\nu_n$ and $g$ from the FDM and CMM can be related from the expressions for the frequency shift induced by the ring-ring coupling
%
\begin{align}\eqlab{model g}
\Phi_1(\omSub{i-}+\Delta\omega) + \arg\{\ssub{\overline{t}}{12}(\omSub{i-}+\Delta\omega)\} \equal 2\pi p ~~\Rightarrow \nonumber\\
\frac{n_gL_1}{c} \Delta\omega + \arctan\!\!\left[\!\frac{-(1-\nu_2^2)\sin\!\Big(\frac{n_gL_2}{c} \Delta\omega\Big)}{2\nu_2 - (1+\nu_2^2)\cos\!\Big(\frac{n_gL_2}{c} \Delta\omega\Big)} \!\right] \equal 2\pi p,
\end{align}
where $\Delta\omega \equal \sqrt{g^2-(\gamSub{i}/4)^2}$.
\figref{model comparison}b plots the transmission calculated from the FDM and CMM using Eqs. (\ref{eq:complex ref index})-(\ref{eq:model g}) to relate the parameters. The good agreement illustrates that the CMM is a good approximation over a fairly large bandwidth close to the resonances.

The effective index, $\tilde{n}_{\rm{eff}}$, group index, $n_g$, and through-coupling, $\ssub{\nu}{n}$, can be calculated from mode solvers such as Lumerical or Comsol given the cross-section geometry of the wave\-guides. Using the CMM and FDM to relate these parameters to estimates of the device performance in~\eqsref{isolation ratio 1}{eta CW max app}, it is possible to design devices with specific properties.

\subsection{\label{app:FDM mzi} Interferometrically Coupled Ring}
To analyze the properties of an interferometrically coupled ring, we introduce a transfer matrix corresponding to~\eqref{coupling matrix} for interferometric coupling
\begin{align}\eqlab{coupling matrix mzi app}
\vsup{C}{(\mathcal{I},n)} = e^{i\ssub{\psi}{R}} \!\!\left[\!\!\!\begin{array}{c c} \ssub{\nu}{n}^2 - e^{i\psi}( 1 - \ssub{\nu}{n}^2) 	& i\ssub{\nu}{n}\sqrt{1\minus\ssub{\nu}{n}^2} \Big(1\plus e^{i\psi}\Big) \\
i\ssub{\nu}{n}\sqrt{1\minus\ssub{\nu}{n}^2} \Big(1\plus e^{i\psi}\Big)  	& \ssub{\nu}{n}^2\Big(1\plus e^{i\psi}\Big) - 1  \end{array} \!\!\!\!\right].
\end{align}
The phase accumulated in the ring (bus waveguide) arm is $\ssub{\psi}{R}$ ($\ssub{\psi}{B}$) and $\psi\equal \ssub{\psi}{B} \minus \ssub{\psi}{R}$. The directional couplers are assumed identical with a through-coupling of $\ssub{\nu}{n}$. Considering a case without input fields ($\ssub{s}{\rm{in}}\equal 0$) and a field, $s_g$, being generated inside ring 1 ($\ssub{s}{1-}\equal \vsubsup{C}{1,1}{(\mathcal{I},1)} \ssub{s}{1+} + s_g$), the solution to~\eqref{field relations} is
\begin{align}\eqlab{intra cavity field IC app}
\frac{\ssub{s}{1-}}{\ssub{s}{g}} =  \frac{1}{1 - \vsubsup{\overline{C}}{1,1}{(\mathcal{I},1)}e^{i\Phi_1}\ssub{\overline{t}}{12} } ,
\end{align}
where $\Phi_1 \equal \phi_1 + \psi_R + \theta_2$ is again the round trip phase of ring 1 without ring 2 present and $\vsup{\overline{C}}{(\mathcal{I},1)}\equal \vsup{C}{(\mathcal{I},1)}e^{-i\psi_R}$. 

\section{Model and Experiment Comparison}\label{app:model experiment comparison}
In this section, we provide additional details on how the model parameters in~\tabref{device params} are estimated. The procedure consists of fitting the measured data using the analytical expressions for transmission and frequency conversion efficiency. We use a step-wise procedure where parameters found from fitting to the transmission spectra without pump fields are used when fitting the transmission and conversion efficiency with the pumps on. The measured transmission spectrum is plotted in~\figref{FDM_experiment}.
\begin{figure}[!h]
  \centering
  \includegraphics[height=4.0cm]{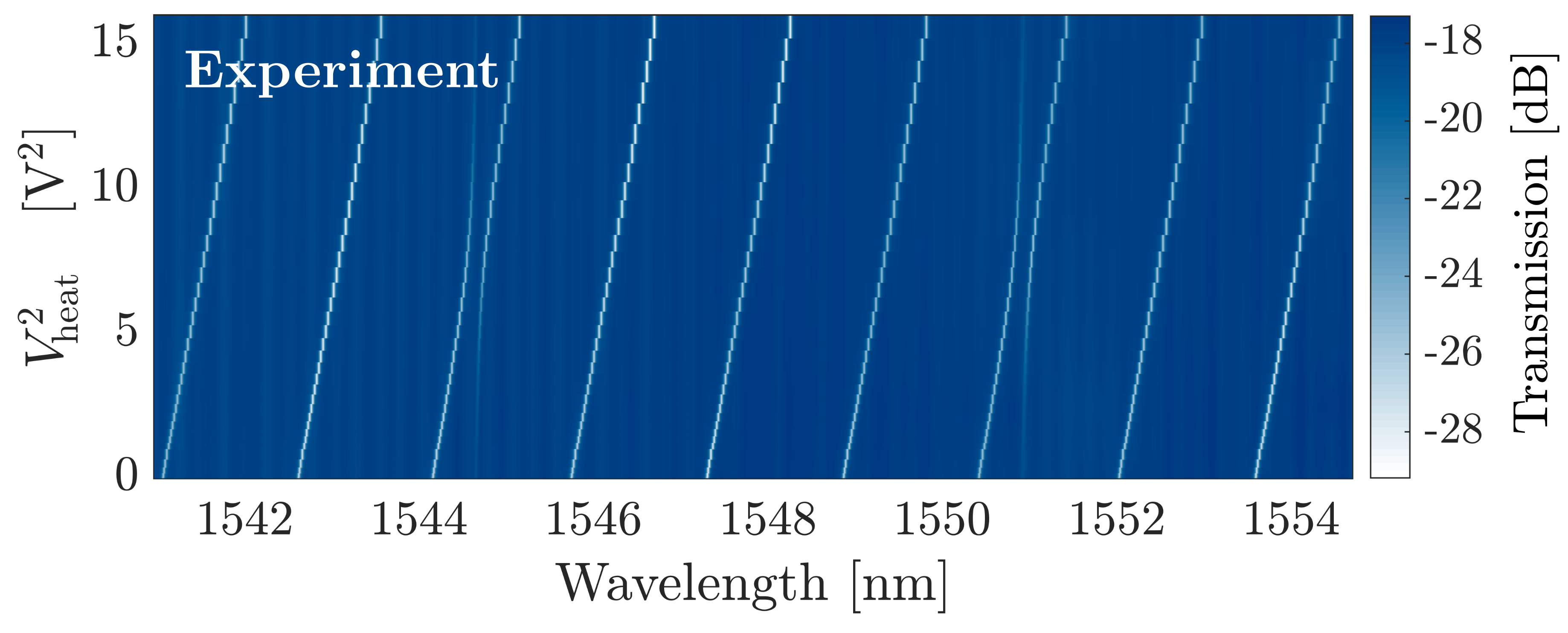}
 \caption{Measured transmission as a function of wavelength and heater voltage. }
\figlab{FDM_experiment}
\end{figure}
To compare to the coupled-mode model (CMM) only parts of the transmission spectrum in the vicinity of the three modes, $\lamSub{i-}$, $\lamSub{s}$, and $\lamSub{i+}$ is used. This is because the CMM is only a good approximation over a bandwidth in which the resonances are well-described by a Lorentzian. The ring-waveguide coupling, $\gamma$, is assumed equal for all modes while ring 1 and ring 2 are allowed different loss rates, $\gamsub{L_1}$ and $\gamsub{L_2}$, respectively. Eqs. (10a) and (10b) then read
\begin{subequations}\eqlab{trans CW CMT app}
\begin{align}\eqlab{trans CW CMT app i-}
T_{i-}(\Omega)  &= T_{\text{cpl}}^2 \left| 1-\frac{\gamSub{}}{ \frac{\gamSub{}+\gamsub{L_1}}{2} - i\Omega + \frac{g^2}{\gamsub{L_2}/2 - i(\Omega-\delta_{ab}) }} \right|^2, \\
T_{i+}(\Omega)  &= T_{\text{cpl}}^2 \left| 1-\frac{\gamSub{}}{    \frac{\gamma + \gamsub{L_1}}{2} - i\Omega } \right|^2 . \eqlab{trans CW CMT app i+}
\end{align}
\end{subequations}
\eqref{trans CW CMT app} includes the coupling loss, $T_{\text{cpl}}$, which is assumed identical for input and output grating couplers.
Changing the heater voltage detunes the resonances of the two rings and we model this by letting $\delta_{ab}$ depend on the voltage as
\begin{align}\eqlab{delta_ab}
\delta_{ab} 	= A + B V_{\rm{heat}}^2 .
\end{align}
Fitting~\eqref{trans CW CMT app} to the data in~\figref{FDM_experiment} allows us to estimate the parameters $\gamma$, $\gamsub{L_1}$, $\gamsub{L_2}$, $g$, $A$, and $B$. The coupling loss, $T_{\text{cpl}}$, is estimated as the mean value of the transmission away from any resonances.
\begin{table}[htbp]
\centering
\begin{tabular}{cc}
\hline
$A$ & $B$  \\  
\hline
$-839 \times 10^{9} \, \mathrm{rad/s}$  & $49.9\times 10^{9} \, \mathrm{rad/s/V^{2}}$ \\
\hline
\end{tabular}
\caption{ Parameters estimated by comparison to transmission measurements without pump lasers. We only list parameters that are not included in~\tabref{device params}.}
  \label{tab:params}
\end{table}
When the pumps are turned on there will be a nonlinear loss in ring 1 due to free-carrier-absorption (FCA) from carriers generated by two-photon-absorption (TPA). This is included in~\eqref{trans CW CMT app} by introducing a modified loss rate only in ring 1, $\overline{\gamma}_{\!L_1}\equal \gamsub{L_1} + \gamsub{\rm{FCA}}$. Additionally, the resonances of ring 1 red-shift due to the heat generated by TPA. This is included by modifying $\delta_{ab}$ as $\Delta_{ab}\equal \delta_{ab} + \delta_{\rm{NL}}$. This leaves only $\bar{\chi}$ to be determined, which is done by also fitting the converted spectra in Figs.~\ref{fig:BSFWM}b-d using modified versions of~\eqsref{isolation ratio 1}{eta CW max app}
\begin{align}\eqlab{isolation ratio}
\zeta(\Omega)  &= \frac{4g^2 }{\big[\gamsub{L_2}+i2(\Delta_{ab}-\Omega)\big]\big(\gamma + \overline{\gamma}_{L_1}-i2\Omega\big)} + 1 \\
\eta_{i+}(\Omega)  &= \left|\frac{4\gamma\bar{\chi} }{\big(\gamma + \overline{\gamma}_{L_1}-i2\Omega\big)^2} \right|^2,~~~~\eta_{i-} = \frac{1}{|\zeta|^2}\eta_{i+} \eqlab{eta i+}.
\end{align}
Since the measured conversion efficiency in our experiment is much smaller than one, we make the simplifying assumption that the signal field is undepleted, which corresponds to neglecting the terms $-i\bar{\chi}^{*}\ssub{A}{i+} -i\bar{\chi}\ssub{A}{i-}$ in~\eqref{EOM as 1}. The parameters $\gamsub{\rm{FCA}}$, $\delta_{\rm{NL}}$, and $\bar{\chi}$ are estimated by simultaneously fitting transmission and idler output power data. Note that the parameters found from transmission data without pumps are held fixed in this process. We use five datasets corresponding to different heater settings, three of which are shown in Figs.~\ref{fig:BSFWM}b-d. The nonlinear loss, $\gamsub{\rm{FCA}}$, and Kerr nonlinearity, $\bar{\chi}$, are assumed to be identical for all five datasets whereas the nonlinear shifts, $\delta_{\rm{NL}}$, are allowed to vary among them. This is due to the fact that the  thermal locking procedure~\cite{Li2016} used to tune the pumps into resonance does not consistently result in exactly the same resonance shift even if the pump power is identical. The values of the nonlinear shifts are listed in~\tabref{shifts} and all five datasets along with the fitted curves are shown in~\figref{BSFWM supp}.\\
\begin{table}[htbp]
\centering
\begin{tabular}{ccccc}
\hline
$\delta^{(1)}_\mathrm{NL}$ & $\delta^{(2)}_\mathrm{NL}$ & $\delta^{(3)}_\mathrm{NL}$ & $\delta^{(4)}_\mathrm{NL}$ & $\delta^{(5)}_\mathrm{NL}$ \\
\hline
$-21.6097$  & $0.8130$  & $4.8078$& $4.1497$ & $11.8393$ \\
\hline
\end{tabular}
\caption{Values of $\delta_\mathrm{NL}$ in units of $10^9 \ \mathrm{rad/s}$ for the five data sets, estimated by fitting transmission and idler power data. We only list parameters not included in~\tabref{device params}. 
}
  \label{tab:shifts}
\end{table}
\begin{figure}[!h]
  \centering
  \includegraphics[height=11.0cm]{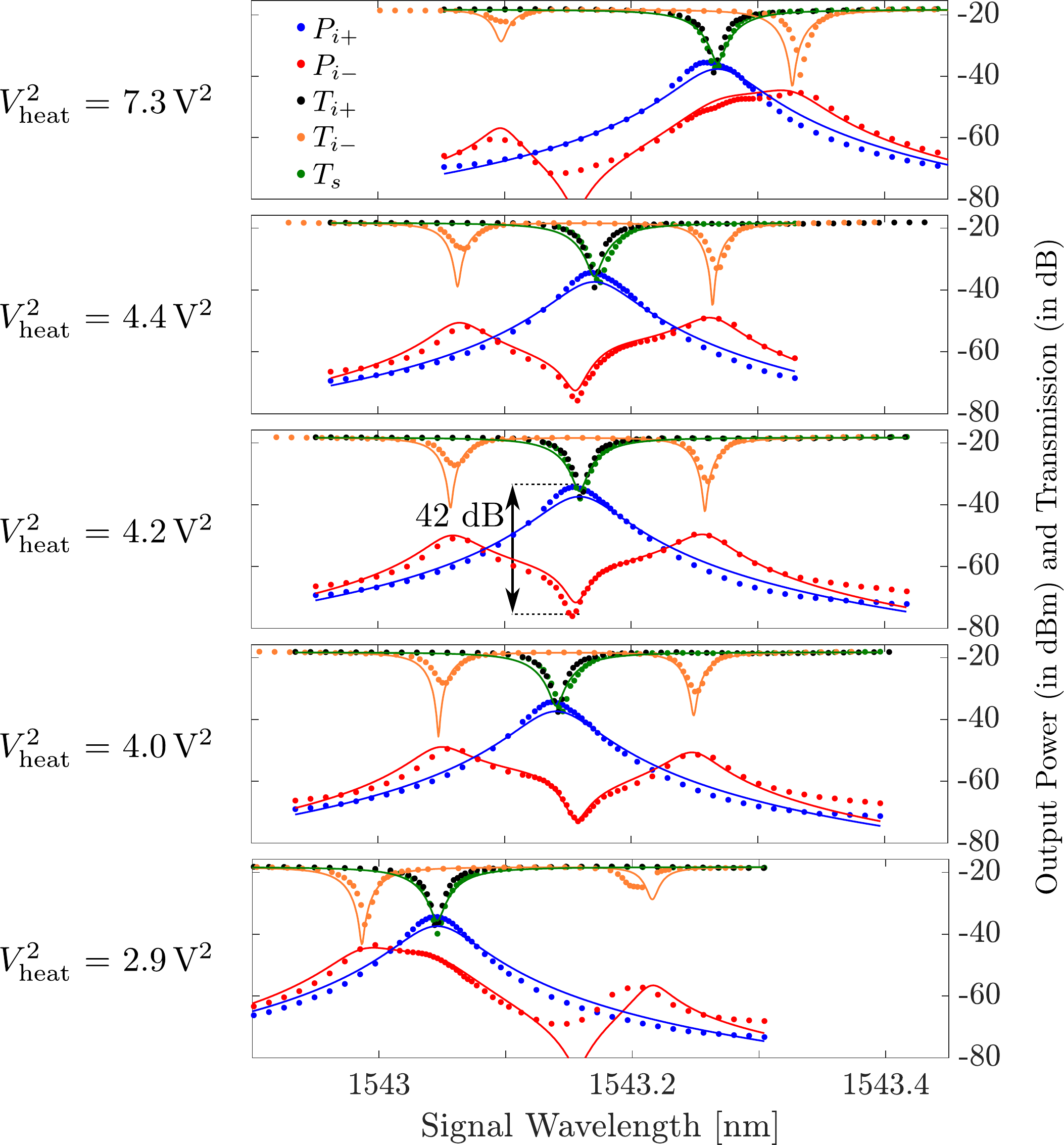}
 \caption{Idler output power as a function of the signal laser wavelength, $\lambda$,  from the down-converted resonance (red) and up-converted resonance (blue) as well as the transmission near $\lamSub{i-}$ (orange), $\lamSub{s}$ (green), and $\lamSub{i+}$ (black). Solid lines are fits and dots are measured data.}
\figlab{BSFWM supp}
\end{figure}

The nonlinearity may also be estimated based on the other parameters as well as the properties of the ring.
Using the results in Ref.~\cite{Vernon2016}, we estimate the nonlinear coupling rate, $\bar{\chi}$, of our device
\begin{align}\eqlab{chi estimate}
|\bar{\chi}| = \Lambda |\overline{\beta}^{*}_{P^{(2)}} \overline{\beta}_{P^{(1)}}| ,
\end{align}
where the parameters on the right hand side are~\cite{Vernon2016}
\begin{align}\eqlab{Lambda estimate}
\Lambda \approx \frac{2\hbar \omref^2 c n_2}{n_{\rm{eff} }'^{2} V_{\rm{ring}}}, \quad |\overline{\beta}_{P^{(j)}}| = 2\frac{\sqrt{\gamSub{j}}}{\GamSub{j}} \sqrt{\frac{P_{p_j}}{\hbar\omref}}.
\end{align}
The nonlinear refractive index is $n_2$ and $V_{\rm{ring}}$ is the volume of ring 1. Assuming both pump modes are identical to the signal mode and inserting~\eqref{Lambda estimate} into~\eqref{chi estimate} yields
\begin{align}\eqlab{chi estimate 2 app}
|\bar{\chi}| =\frac{8  \omref c n_2}{n_{\rm{eff} }'^{2} V_{\rm{ring}}}\frac{\gamSub{s}}{\GamSub{s}^2}\sqrt{P_{\!p_1} P_{\!p_2}}.
\end{align}
%

\subsection{Fitting using Frequency Domain Model}
As an additional check of the fitted parameters using the CMM, we also use the FDM to fit the spectra in~\figref{FDM_experiment}. Here, we use the entire wavelength range of the measurement as the FDM models all the resonances as well as the FSR of the rings.~\eqref{trans t} is used to estimate the parameters $\nu_1$, $\nu_2$, $\tilde{n}_{\rm{eff}}$, and $n_g$. We assume only the real part of $\tilde{n}_{\rm{eff}}$ changes with applied voltage as in~\eqref{delta_ab}
\begin{align}\eqlab{thermal model}
n_{\rm{eff}}'(V_{\rm{heat}}) = n_0' +  \partial n_V V_{\rm{heat}}^2.
\end{align}
The proportionality constant is
\begin{align}\eqlab{thermal model 2}
\partial n_V = \frac{1}{R}\frac{\partial n}{\partial T}  \frac{\partial T}{\partial U}  
\end{align}
where $R$ is the resistance of the titanium wire, $\partial U/\partial T$ is its heat capacity, and $\partial n/\partial T$ is the thermo-optic coefficient of silicon.~\figref{FDM_model}a shows a comparison with~\figref{FDM_experiment} using the fitting parameters listed in~\tabref{FDM device params}. 
\begin{figure}[!h]
  \centering
  \includegraphics[height=4.0cm]{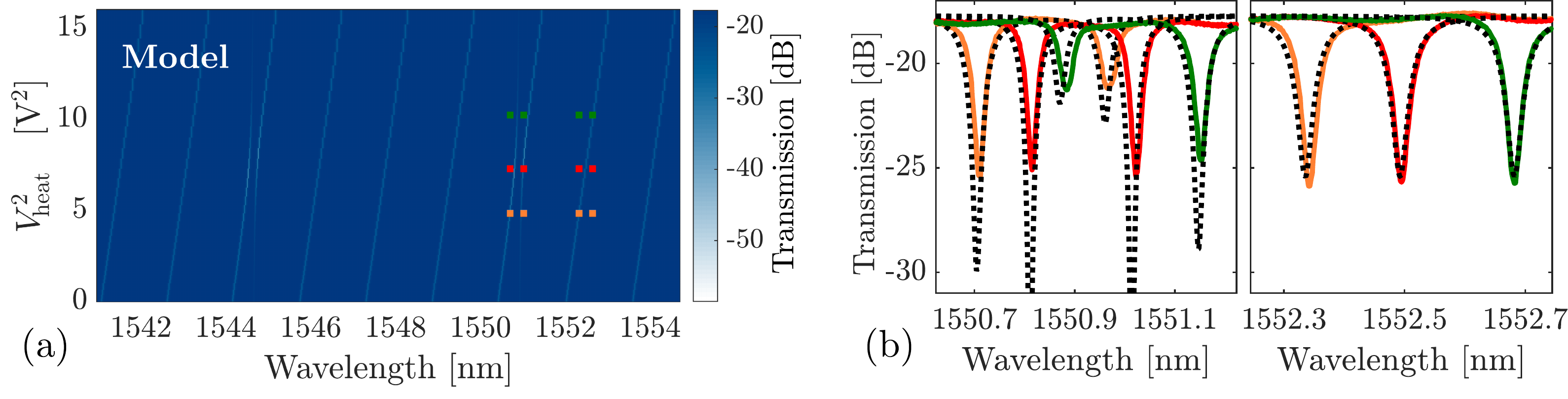}
 \caption{(a) Plot of~\eqref{trans t} for the transmission as a function of wavelength and heater voltage. (b) Spectra from the cross-sections indicated by dotted lines in (a). }
\figlab{FDM_model}
\end{figure}
The largest discrepancy is observed for heater voltages where the resonances of the rings align. 
\begin{table}[htbp]
\centering
\caption{\bf FDM fitting parameters.}
\begin{tabular}{p{0.3cm} p{0.05cm}  p{2.5cm}  p{0.05cm} | p{0.05cm}   p{0.3cm}  p{0.05cm} p{2.5cm}  }
\hline
$\gamma$  		&=& $28\!\times\!10^9\,\text{rad/s}$ 	& && $\gamma_{L_1}$  &=& $12\!\times\!10^9\,\text{rad/s}$ \\
$g$  			&=& $80\!\times\!10^9\,\text{rad/s}$ 	& && $\gamma_{L_2}$  &=& $23\!\times\!10^9\,\text{rad/s}$ \\
$n_0'$  		&=& $2.618$ 				& && $n_g$  	&=& $4.73$ 				\\
$\partial n_V$  &=& $1.96\!\times\!10^{-4}$ V$^{-2}$ 	& && $ $  		& & $ $ 	\\
\hline
\end{tabular}
  \label{tab:FDM device params}
\end{table}
However, when ring 1 is either red- or blue-detuned from ring 2, the agreement is better as observed in~\figref{FDM_model}b. The model agrees well with the measurement data at frequencies close to modes of ring 1 that are far detuned from modes of ring 2, which is also observed from~\figref{FDM_model}b. The values of parameters in~\tabref{FDM device params} agree well with those in~\tabref{device params}, which provides additional confidence in our parameter estimation.

\section{Cascaded FWM}\label{app:cascaded FWM}
To consider the effect of first-order cascaded BS-FWM processes we introduce an extra mode $\ssub{A}{i++}$ that couples to $\ssub{A}{i+}$
\begin{subequations}\eqlab{cascaded EOM}
\begin{align}
	\ssub{\dot{A}}{s} 	&= - \frac{\GamSub{s}}{2} \ssub{A}{s} -i\bar{\chi}^{*}\ssub{A}{i+} -i\bar{\chi}\ssub{A}{i-} - \sqrt{\gamSub{s}}\Si{s} \eqlab{EOM as}\\
	\ssub{\dot{A}}{i+}	&= - \frac{\GamSub{i}}{2}\ssub{A}{i+} - i\bar{\chi} \ssub{A}{s} -i\bar{\chi}^{*}\ssub{A}{i++} \eqlab{EOM ai+}\\	
	\ssub{\dot{A}}{i++}	&= - \frac{\GamSub{i}}{2}\ssub{A}{i++} - i\bar{\chi} \ssub{A}{i+}   \eqlab{EOM ai++}\\		
    \ssub{\dot{A}}{i-} 	&=  - \frac{\GamSub{i}}{2}\ssub{A}{i-} - i\bar{\chi}^{*}\ssub{A}{s} -ig\ssub{B}{}  \eqlab{EOM ai-} \\ 
    \ssub{\dot{B}}{} 	&= \left(-i\delSub{ab} - \frac{\gamSub{L}}{2} \right)\ssub{B}{} -ig\ssub{A}{i-} \eqlab{EOM b}\\
    \So{j} &= \Si{j} + \sqrt{\gamSub{j}}\ssub{A}{j}, \quad j \!\in\! \{s, i-, i+, i\!+\!+\}.
\end{align} 
\end{subequations}
Again, we have assumed that all idler modes have the same coupling rates. For $\delSub{ab}\equal 0$ and $\Omega\equal 0$, we find the extinction ratio
\begin{align}\eqlab{isolation ratio cascaded}
\zeta =   \frac{ \GamSub{i}^2\big(4G^2+1\big)}{ \GamSub{i}^2+ 4\bar{\chi}^2 }  .
\end{align}
The up-conversion efficiency is
\begin{align}\eqlab{eta CW cascaded}
	\eta_{i+} &\!=\! \left| \frac{4\sqrt{\gamSub{i}\gamSub{s}}\bar{\chi}}{\GamSub{i}\GamSub{s} + 4\bar{\chi}^2\left(1+\frac{\GamSub{s}}{\GamSub{i}}+ \zeta^{-1} \right)}  \right|^2 .
\end{align} 
For $\zeta\gg 1$, the maximum conversion efficiency is
\begin{align}\eqlab{eta CW max cascaded}
\eta_{i+}^{\rm{max}} =  \frac{\gamSub{i}\gamSub{s}}{\GamSub{s}(\GamSub{i}+\GamSub{s})} 
\end{align} 
which is achieved with a nonlinearity given by
\begin{align}\eqlab{chi max cascaded}
\chi^{\rm{max}} =  \frac12 \GamSub{i}\sqrt{\frac{\GamSub{s}}{\GamSub{i}+\GamSub{s}}}.
\end{align} 
Inserting~\eqref{chi max cascaded} into~\eqref{isolation ratio cascaded} yields
\begin{align}\eqlab{isolation ratio cascaded max}
\zeta^{\text{max}} =  \big(4G^2+1\big) \frac{ \GamSub{i}+\GamSub{s}}{ \GamSub{i}+2\GamSub{s} }  .
\end{align}
From Eqs. (\ref{eq:isolation ratio cascaded})-(\ref{eq:isolation ratio cascaded max}) it is seen that the additional requirement $\GamSub{i}\gg\GamSub{s}$ must be imposed to reach near-unity conversion efficiency when taking cascaded processes into account. In fact,~\eqref{eta CW max cascaded} shows that the conversion efficiency is limited to 50\% if $\gamSub{i}\equal\gamSub{s}$. This number would be even lower when considering second- and higher-order cascaded processes.
\section{Spectral Correlations}\label{app:pair production app}
The joint state of a photon pair created at $\omSub{s,0}$ and $\omSub{i,0}$ after they exit the resonator is
\begin{equation}
|\Psi \rangle = \iint \mathrm{d}\Omega_i \mathrm{d}\Omega_s \mathcal{A}(\Omega_s,\Omega_i) \hat{\phi}^\dagger_s(\Omega_s)\hat{s}^\dagger_i(\Omega_i),
\end{equation}
where $\hat{s}^\dagger_i(\OmSub{i})$ creates a photon in the waveguide at the frequency $\omega\equal \omSub{i,0}\plus \Omega_i$. Since $\gamSub{s}\approx 0$, the signal photon only couples to the environment and $\hat{\phi}^\dagger_s(\OmSub{s})$ is therefore the creation operator for heat bath modes at $\omega\equal \omSub{s,0}\plus \Omega_s$.
The joint spectral amplitude (JSA), $\mathcal{A}(\Omega_s,\Omega_i)$, is essentially a two-dimensional wave function containing information about the distribution and correlations of the signal and idler frequencies. For photons created in a resonator, the JSA is proportional to \cite{Helt2010,Vernon2017}
\begin{equation}
\label{eq:state}
\mathcal{A}(\Omega_s,\Omega_i) = F_p(\Omega_s+\Omega_i)l_i(\Omega_i)l_s(\Omega_s) ,
\end{equation}
where $l_j(\OmSub{j}) = [\Gamma_j/2+i\OmSub{j}]^{-1}$ are Lorentzian lineshapes with a width determined by $\Gamma_j$ with $j\!\in\!\{s,i,p\}$. The pump function $F_p$ is given by
\begin{equation}\eqlab{Fp}
F_p(\Omega) = \int \mathrm{d}\Omega' \ssub{A}{p}(\Omega-\Omega') l_p(\Omega-\Omega')\ssub{A}{p}(\Omega')l_p(\Omega'),
\end{equation}
which is a convolution of the pump field in the ring, $\ssub{A}{p}(\Omega) l_p(\Omega)$, with itself. The JSA may be expanded using a Schmidt decomposition
\begin{equation}
\mathcal{A}(\Omega_s,\Omega_i) = \sum_k \lambda_k \psi_{i,k}(\Omega_i) \psi_{s,k}(\Omega_s),
\end{equation}
where $\psi_{j,k}$, $j\!\in\!\{s,i\}$ are orthonormal signal and idler Schmidt modes and $\lambda_k$ the Schmidt coefficients. The square of the signal density matrix (originating from a partial trace of the joint state operator over the idler sub-space) is used to define the spectral purity of heralded photons~\cite{Christensen2018}
\begin{equation}
P = \sum_k \lambda_k^4.
\end{equation}
All information about signal-idler frequency correlations are contained in $F_p(\OmSub{s}+\OmSub{i})$ as seen from~\eqref{state}. The maximum spectral purity is achieved when the spectral width of the pump pulse is much larger than the width of the pump mode and the approximation $A_p\approx 1$ can be made in~\eqref{Fp}. 

\section{Shaping Output Photons}\label{app:Shaping Output Photons}
The temporal shape of emitted photons is controllable via the time dependent BS-FWM coupling terms in~\eqref{EOM 1}. To determine the function, $\bar{\chi}_{\rm{in}}(t)$, giving rise to a specific output, we consider the equations of motion for an interferometrically coupled device 
\begin{subequations}\eqlab{eom no cross phase}
\begin{align}
	\ssub{\dot{A}}{o} 	&= - \frac{\GamSub{o}}{2} \ssub{A}{o} -i\bar{\chi}_{\rm{in}}^{*}\ssub{A}{s}  - \sqrt{\gamSub{o}}\Si{o} \eqlab{EOM ncp as}\\
	\ssub{\dot{A}}{s} 	&= -\frac{\gamSub{L}}{2}\ssub{A}{s} - i\bar{\chi}_{\rm{in}} \ssub{A}{o} \eqlab{EOM ncp ai+}\\	
    \So{o} &= \Si{o} + \sqrt{\gamSub{o}}\ssub{A}{o}. \eqlab{EOM ncp IO}
\end{align} 
\end{subequations}
For simplicity we consider only two modes (corresponding to the limit of large $G$), the signal ($s$) and output ($o$). Instead of considering how to emit a function $\So{o}(t)$ in the absence of any inputs ($\Si{o}\equal 0$), we consider how to absorb a function $\Si{o}(-t)\equal \So{o}(t)$. If $\bar{\chi}_{\rm{in}}(t)$ is the control function that enables absorption of $\Si{o}(-t)$, then $\bar{\chi}_{\rm{out}}(t)\equal \bar{\chi}_{\rm{in}}(-t)$ is the control function that enables emission of $\So{o}(t)$ (in the limit of zero loss). Determining $\bar{\chi}_{\rm{in}}(t)$ therefore solves both the absorption and emission problem.

To fully absorb a pulse into the resonator, we must have $\So{o}\equal 0$ in~\eqref{EOM ncp IO} and therefore $\ssub{A}{o}\equal -\Si{o}/\sqrt{\gamSub{o}}$. Inserting into~\eqref{EOM ncp as} yields
\begin{align}\eqlab{EOM ncp as 2}
	\dotSi{o} 	= -\frac{\GamSub{o}}{2} \Si{o} + i\bar{\chi}^{*}_{\rm{in}}\sqrt{\gamSub{o}}\ssub{A}{s}  + \gamSub{o}\Si{o} \quad \Rightarrow \quad 
	\dotSi{o} -\frac{\gamSub{o}-\gamSub{L}}{2} \Si{o}  	=  i\bar{\chi}^{*}_{\rm{in}}\sqrt{\gamSub{o}}\ssub{A}{s}. 
\end{align} 
The solution for $\ssub{A}{s}$ is found by rearranging terms in~\eqref{EOM ncp ai+}
\begin{align} \eqlab{EOM ncp ai+ 2}
    \frac{d}{dt}\!\Big( \ssub{A}{s} e^{\frac{\gamSub{L}}{2} t} \Big) e^{-\frac{\gamSub{L}}{2} t}  	= i\frac{\bar{\chi}_{\rm{in}}}{\sqrt{\gamSub{o}}}\Si{o} \quad \Rightarrow \quad  \ssub{A}{s}  = \frac{i e^{-\frac{\gamSub{L}}{2} t}}{\sqrt{\gamSub{o}}} \int_0^t e^{\frac{\gamSub{L}}{2} t'} \bar{\chi}_{\rm{in}}(t')\Si{o}(t') dt'. 
\end{align} 
Inserting the result for $\ssub{A}{s}$ into~\eqref{EOM ncp as 2} we find
\begin{align}\eqlab{EOM ncp as 3}
	\Big(\frac{\gamSub{o}-\gamSub{L}}{2} \Si{o} - \dotSi{o} \Big)\Si{o} e^{\gamSub{L} t} = \bar{\chi}^{*}_{\rm{in}} e^{\frac{\gamSub{L}}{2} t}\Si{o} \int_0^t e^{\frac{\gamSub{L}}{2} t'} \bar{\chi}_{\rm{in}}(t')\Si{o}(t') dt'.
\end{align} 
We assume $\Si{o}\!\in\!\mathbb{R}$ such that the RHS can be written as
\begin{align}\eqlab{complex product}
	 (x-iy) \int\! (x+iy) =  x\int\!x + y\int\! y + i\Big(x\int\! y - y\int\! x\Big),
\end{align} 
where the functions are given by $x\equal \text{Re}\{\bar{\chi}_{\rm{in}}\}\Si{o}\exp(\gamSub{L}t/2)$ and similarly $y\equal \text{Im}\{\bar{\chi}_{\rm{in}}\}\Si{o}\exp(\gamSub{L}t/2)$. By defining the functions
\begin{align}\eqlab{aux functions}
	 X=\int\!x = R\cos(\theta), ~ Y=\int\!y = R\sin(\theta),
\end{align} 
the real part of~\eqref{complex product} can be rewritten as 
\begin{multline}\eqlab{RHS real}
	 \dot{X}X + \dot{Y}Y = \big[\dot{R}\cos(\theta) - R\sin(\theta)\dot{\theta}\big]R\cos(\theta) + \big[\dot{R}\sin(\theta) + R\cos(\theta)\dot{\theta} \big]R\sin(\theta) =~\\ 
	\dot{R}R = \frac12 \frac{d}{dt}\!\Big( R^2\Big).
\end{multline} 
The imaginary part of~\eqref{complex product} is 
\begin{align}\eqlab{RHS imag}
	 \dot{X}Y - \dot{Y}X = \big[\dot{R}\cos(\theta) - R\sin(\theta)\dot{\theta}\big]R\sin(\theta) - \big[\dot{R}\sin(\theta) + R\cos(\theta)\dot{\theta} \big]R\cos(\theta) = -R^2\dot{\theta}.
\end{align} 
Since $\Si{o}$ is real, we can define a real function 
\begin{align}\eqlab{fin definition}
	 f_o = \Big(\frac{\gamSub{o}-\gamSub{L}}{2} \Si{o} - \dotSi{o} \Big)\Si{o} e^{\gamSub{L} t},
\end{align} 
such that the solution to~\eqsref{RHS real}{RHS imag} are
%
\begin{align}\eqlab{mcR solution}
	 R(t) 		&= \sqrt{2 \int_0^t\! f_o(s) ds} \\
     \dot{\theta}&= 0 \quad \Rightarrow \quad \theta = C_1,
\end{align} 
where $C_1$ is an arbitrary real constant. For simplicity we could choose $C_1\equal 0$, such that $X\equal R$ and therefore
\begin{align} \eqlab{chi sol}
	 \bar{\chi}_{\rm{in}} \Si{o}e^{\frac{\gamSub{L}}{2} t} = \dot{R} = \frac{f_o}{\sqrt{2 \int\! f_o}}  \quad \Rightarrow \quad  \bar{\chi}_{\rm{in}} =  \frac{f_o e^{-\frac{\gamSub{L}}{2} t}}{ \Si{o} \sqrt{2 \int\! f_o}} . 
\end{align} 
%
\subsection{Gaussian wave packet}\label{app:Gaussian wave packet}
Having determined the general solution in~\eqref{chi sol}, we now consider a specific example of a Gaussian wave packet
\begin{align}\eqlab{Gaussian}
\Si{o}(t) =  \sqrt{\frac{2}{\Delta t}} \left(\frac{\text{ln}(2)}{\pi}\right)^{\!\frac{1}{4}} \exp\!\left(\!-2\text{ln}(2)\frac{t^2}{\Delta t^2} \right) ,
\end{align} 
with a full width at half maximum (FWHM) temporal width $\Delta t$ and spectral width $\Delta\omega\equal 4\text{ln}(2)/\Delta t$. First, we note that~\eqref{chi sol} is only a solution if the anti-derivative of $f_o$ is positive. For the Gaussian wave packet in~\eqref{Gaussian} we have 
\begin{align}\eqlab{Gaussian derivative}
\dotSi{o} =  -\frac{4\text{ln}(2)}{\Delta t^2} t \Si{o} = -\frac{\Delta\omega}{\Delta t} t \Si{o}.
\end{align} 
Since $\Si{o}$ is positive, the condition for $f_o$ to be positive is
\begin{align} \eqlab{f_o positive}
f_o\geq 0 \quad \Rightarrow \quad \frac{\gamSub{o}-\gamSub{L}}{2} + \frac{\Delta\omega}{\Delta t} t \geq 0 \quad \Rightarrow \quad \frac{t}{\Delta t} \geq -\frac12 \frac{\gamSub{o}}{\Delta\omega}\Big(1-\frac{\gamSub{L}}{\gamSub{o}}\Big) . 
\end{align} 
This means $f_o$ is negative in the leading part of the wave packet up until the time given by~\eqref{f_o positive} and the solution for $\bar{\chi}_{\rm{in}}$ is invalid. However, this critical time can be pushed arbitrarily far into the tail of the Gaussian by increasing the coupling rate of the output mode, $\gamSub{o}$, relative to the pulse bandwidth, $\Delta\omega$. 

To calculate the emission efficiency in the presence of loss and the down-converted mode, we consider the equations of motion
\begin{subequations}\eqlab{eom emission}
\begin{align}
	\ssub{\dot{A}}{o} 	&= - \frac{\GamSub{o}}{2} \ssub{A}{o} -i\bar{\chi}_{\rm{out}}\ssub{A}{s}  \eqlab{eom emission Ao}\\
	\ssub{\dot{A}}{s} 	&= -\frac{\gamSub{L}}{2}\ssub{A}{s} - i\bar{\chi}_{\rm{out}} \ssub{A}{o} - i\bar{\chi}_{\rm{out}} \ssub{A}{i-} \eqlab{eom emission As}\\	
	\ssub{\dot{A}}{i-} 	&= -\frac{\GamSub{o}}{2}\ssub{A}{i-} - i\bar{\chi}_{\rm{out}} \ssub{A}{s} - i g \ssub{B}{} \eqlab{eom emission Ai-}\\
    \ssub{\dot{B}}{}	&= \left(-i\delSub{ab} - \frac{\gamSub{L}}{2} \right)\ssub{B}{} -ig\ssub{A}{i-} \eqlab{eom emission B} \\
    \So{o} &= \sqrt{\gamSub{o}}\ssub{A}{o},
\end{align} 
\end{subequations}
For simplicity, we assume the down-converted mode and output mode have the same coupling rates. The appropriate initial condition is $\ssub{A}{s}(0)\equal 1$, $\ssub{A}{o}(0)\equal\ssub{A}{i-}(0)\equal\ssub{B}{}(0)\equal 0$, and $\bar{\chi}_{\rm{out}}(0)\equal 0$. This corresponds to a signal photon occupying mode $\ssub{A}{s}$ and the BS-FWM pumps not having entered the resonator yet. The absolute square of the cavity fields then correspond to the probability of the photon occupying these modes, $\mathcal{P}(j)\equal |\ssub{A}{j}|^2$. As mentioned above, the solution for $\bar{\chi}_{\rm{out}}(t)$ is given by~\eqref{chi sol} evaluated at $-t$. It turns out to be advantageous to multiply $\bar{\chi}_{\rm{out}}$ by an envelope function, $F_{\rm{env}}(t)\equal f_{\rm{env}}(t\minus\tau_{\rm{on}})\!\times\! f_{\rm{env}}(\tau_{\rm{off}}\minus t)$, due to loss and the divergence in $\bar{\chi}_{\rm{in}}$. The sides of the envelope are
\begin{align}\eqlab{envelope function main}
	f_{\text{env}}(t) = \bigg[1\plus \frac{1 \plus \sin\!\big( \frac{\pi t}{\tau_{\rm{env}}}\big) }{2}  \theta\Big(t \plus \frac{\tau_{\rm{env}}}{2}    \Big)  \bigg]  \theta\Big(t\minus \frac{\tau_{\rm{env}}}{2}   \Big) ,
\end{align} 
where $\theta(t)$ is a step function that equals one when $t\!>\!0$ and zero otherwise. The envelope rises from zero to one in the interval $t\!\in\![\minus \tau_{\rm{env}}/2,~ \tau_{\rm{env}}/2]$ as half a period of the sine function. 
\begin{figure}[!h]
  \centering
  \includegraphics[height=3.3cm]{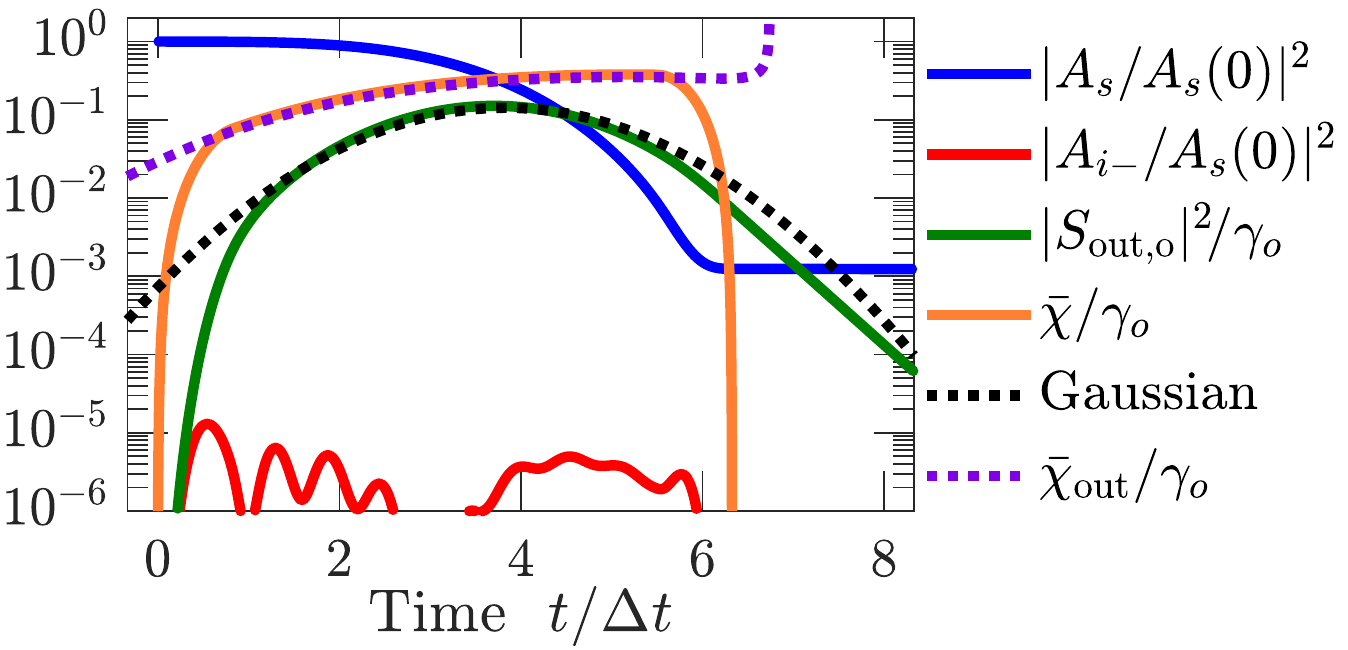}
 \caption{Example of the emitted wave packet calculated from~\eqref{eom emission}. The Gaussian (dotted black) plots~\eqref{Gaussian} with an appropriate temporal shift. The used parameters are: $Q_L/Q_o\equal 500$, $G\equal 100$, $\Delta\omega/\gamSub{o}\equal 0.38$, $t_0\equal 3.83\Delta t$, $\tau_{\rm{on}}\equal 0.42\Delta t$, $\tau_{\rm{off}}\equal 5.91\Delta t$, and $\beta\equal 1.06$. }
\figlab{Gaussian_Emission example app}
\end{figure}
~\figref{Gaussian_Emission example app} shows an example of the solution to~\eqref{eom emission} along with the BS-FWM pump function given by $\bar{\chi}\equal  \beta \bar{\chi}_{\rm{out}} F_{\rm{env}}$. The amplitude, $\beta\!>\! 1$, is used to ensure that the entire population in $\ssub{A}{s}$ may be converted to $\ssub{A}{o}$ despite the reduction in area under the orange curve resulting from multiplication by $F_{\rm{env}}$. The function $\bar{\chi}$ (orange curve in~\figref{Gaussian_Emission example app}) is found by optimizing $\tau_{\rm{on}}$, $\beta$, and $\Delta\omega/\gamSub{o}$ to maximize $\eta_{\rm{out}}$ for a fixed $G$ and $Q_L/Q_o$ under the constraint that $\text{OL}\geq 99\%$. The cutoff time, $\tau_{\rm{off}}$, is fixed for a given pulse width, $\Delta\omega$, to avoid the divergence in $\bar{\chi}_{\rm{out}}$ (occurring at $t\!\approx\! 7\Delta t$ in~\figref{Gaussian_Emission example app}).

\section*{References}

\bibliographystyle{ieeetr}

\end{document}